\documentclass[10pt,twocolumn,twoside]{IEEEtran}
\usepackage{amsmath}
\usepackage{amssymb}
\usepackage{algorithmicx}
\usepackage{algpseudocode}
\usepackage{graphicx}
\usepackage{subfigure}
\usepackage{epstopdf}
\usepackage{cite}
\usepackage{textcomp}
\usepackage{mathrsfs}
\usepackage{color}
\usepackage{booktabs}
\usepackage{cases}
\usepackage{setspace}
\usepackage{bm}
\usepackage{cuted}
\usepackage{booktabs}
\usepackage{stfloats}
\usepackage{makecell}
\usepackage{mathtools}
\usepackage{arydshln}
\usepackage{bbding}
\usepackage[ruled,linesnumbered]{algorithm2e}

\SetKwRepeat{Do}{do}{while}
\makeatletter

\renewcommand{\maketag@@@}[1]{\hbox{\m@th\normalsize\normalfont#1}}%

\makeatother

\begin{document}
	
	\title{\LARGE Exploring Hybrid Active and Passive Multiple Access via Slotted ALOHA-Driven Backscatter Communications} 
	
		\author{
      Bowen~Gu,  Dong~Li, Hao~Xie, Kan Yu, Quansheng Guan, Yongjun~Xu

					\vspace{-1em}
					\IEEEcompsocitemizethanks{ 
						\IEEEcompsocthanksitem This work was supported in part by Science and Technology Development Fund, Macau SAR, under Grant 0188/2023/RIA3,  in part by  the National Natural Science Foundation of China with Grant 62301076, and in part by the Macao Young Scholars Program with Grant AM2023015.
						\IEEEcompsocthanksitem Bowen Gu, Dong Li, and Hao Xie are with the School of Computer Science and Engineering, Macau University of Science and Technology, Avenida Wai Long, Taipa, Macau 999078, China (e-mails: gubwww@163.com  dli@must.edu.mo, 3220005631@student.must.edu.mo) (\textit{Corresponding author: Dong Li}).  
						\IEEEcompsocthanksitem Kan Yu is with the School of Computer Science and Engineering, Macau University of Science and Technology, Avenida Wai Long, Taipa, Macau 999078, China, and  also the Key Laboratory of Universal Wireless Communications
						Ministry of Education,
						Beijing University of Posts and Telecommunications
						Beijing 100876, China China (e-mail: 
						kanyu1108@126.com).
						\IEEEcompsocthanksitem Quansheng Guan is with the School of
						Electronics and Information Engineering, South China University of
						Technology, Guangzhou 510641, China (e-mail: eeqshguan@scut.edu.cn).
						\IEEEcompsocthanksitem Yongjun Xu is with the School of Communication and Information Engineering, Chongqing University of Posts and Telecommunications, Chongqing 400065, China (e-mails: xuyj@cqupt.edu.cn).
					}
		}

	%

	\maketitle
	\thispagestyle{empty}
	\pagestyle{empty}
	
	\begin{abstract}		
         In conventional backscatter communication (BackCom) systems, various centralized access schemes, i.e., time/frequency division multiple access, are commonly adopted to prevent transmission collisions among multiple backscatter devices (BDs). However, as the number of BDs proliferates, there will be high overhead, and more seriously, the proactive inter-user coordination may be unaffordable for the passive BDs, which are of scarce concern in existing works and remain a challenge to solve. To this end, in this paper, we propose a slotted ALOHA-based random access for multi-BD BackCom systems, where each BD is selected randomly and coexists with an active device for hybrid transmission. To excavate and evaluate the performance, a resource allocation problem targeting max-min average throughput is formulated, where transmit antenna selection,  receive beamforming design,  reflection coefficient (RC)  adjustment, power control, and access probability determination are jointly optimized. To deal with this intractable problem,  we first transform the  objective function with the max-min form into an equivalent linear one, and then decompose the resulting problem into three sub-problems. Next, a block coordinate descent-based greedy algorithm, incorporating maximum ratio combining, semi-definite programming, and variable substitution, to achieve a sub-optimal solution. Besides, we derive closed-form solutions for both RC and transmit power to facilitate analysis.  Simulation results demonstrate that the proposed algorithm outperforms benchmark algorithms in terms of throughput and fairness.
	\end{abstract}
	
	\IEEEpeerreviewmaketitle
	\section{Introduction}
As massive wireless devices are continuously connected to the Internet, the development of human society has transitioned from the period of information revolution, epitomized by the Internet, to the period of intelligent revolution, ushered in by the Internet of Things (IoT), heralding the opening of the door to the Internet of Everything \cite{Vaezi,LiuWC,LiuLiDai}. According to forecasts, by 2027, the number of active IoT devices will skyrocket to 29.7 billion, contributing significantly to a more interconnected world \cite{Centenaro}. 	However, a significant concern emerges regarding the energy supply for these substantial numbers of IoT devices, given the current energy scarcity and considerable energy consumption of each device \cite{XuGuHuLi}. Consequently, reducing the energy consumption of IoT devices has become a matter of urgency.

Backscatter communication (BackCom), known for its cost and energy efficiency, holds the promise of mitigating the power consumption challenges inherent to these devices \cite{NiuLiY}. Precisely, IoT devices can adjust the reflection coefficient (RC) by changing the impedance, thus reflecting and modulating the incident radio-frequency (RF) signal from a dedicated signal source without the carrier signal generated by oscillators. As a result, BackCom bypasses the need for demodulation and decoding and eliminates the requirement for additional RF components. This makes its circuit power consumption significantly lower than conventional communication methods. Along with the evolution of BackCom, ambient backscatter communication (AmBC) inherits the hallmarks of BackCom, and differs from the former in that AmBC is allowed to utilize the surrounding RF sources or systems like cellular base stations (BSs), digital TV transmitters, Wi-Fi access points, frequency modulation (FM), LoRa, and Bluetooth signals  \cite{Huynh}.
	
	\subsection{Related Works}	
%

\begin{table*}[t]
	\vspace{-3mm}
	\newcommand{\tabincell}[2]{\begin{tabular}{@{}#1@{}}#2\end{tabular}}
\scriptsize
	\centering
	\label{t1}
	\caption{Summary of the Related Literature for BackComs with Resource allocation.}	
	\begin{tabular}{ccccccccccc}
		\hline
		Ref. & \tabincell{c}{Network \\type} & \tabincell{c}{BD \\number}&  \tabincell{c}{Multiple \\access} &\tabincell{c}{Cooperative\\transmission}&\tabincell{c}{Beamforming\\ Design} &  \tabincell{c}{ Fairness }& \tabincell{c}{Non-linear\\EH}& \tabincell{c}{DLI\\ cancellation} & \tabincell{c}{ MI \\cancellation} & \tabincell{c}{Random \\control}    \\
		
		\hline
		[9] & \tabincell{c}{SISO} & Single& \tabincell{c}{N/A}  & N/A  &  \tabincell{c}{N/A} &	N/A  & 
	\tabincell{c}{\XSolidBrush} & 
		\tabincell{c}{ \CheckmarkBold} (SIC)  &  \tabincell{c}{N/A} &  
		\tabincell{c}{\XSolidBrush}    \\
		
		\hline
		
		[10]& \tabincell{c}{MIMO} & Multiple& \tabincell{c}{TDMA} &N/A  & \tabincell{c}{ \CheckmarkBold}& \tabincell{c}{\XSolidBrush} &
\tabincell{c}{ \CheckmarkBold} & \tabincell{c}{\CheckmarkBold}  (SIC)& 
		\tabincell{c}{ \CheckmarkBold} (TDMA) &\tabincell{c}{\XSolidBrush}  \\
		
		\hline
		
		[11] & \tabincell{c}{SISO} &Multiple&  \tabincell{l}{TDMA}  & N/A  & N/A & \tabincell{c}{\CheckmarkBold}  & \tabincell{c}{ \CheckmarkBold} & \tabincell{c}{\CheckmarkBold} (SIC)& 
		\tabincell{c}{ \CheckmarkBold}(TDMA)  &\tabincell{c}{\XSolidBrush}   \\
		
		\hline

		[12] & \tabincell{c}{SISO} & Multiple&  \tabincell{l}{TDMA}  & N/A  & N/A &  \tabincell{c}{\CheckmarkBold}  & 	\tabincell{c}{\XSolidBrush} &  \tabincell{c}{\XSolidBrush} & 
		\tabincell{c}{ \CheckmarkBold} (TDMA) & \tabincell{c}{\XSolidBrush}  \\

		\hline
		
		[13] & \tabincell{c}{SISO} & Multiple&  \tabincell{c}{FDMA}  & N/A  & N/A & \tabincell{c}{\XSolidBrush} & N/A& \tabincell{c}{\CheckmarkBold} (FS) &\tabincell{c}{\CheckmarkBold} (FS)&  \tabincell{c}{\XSolidBrush} \\
		
		\hline
		
		[14] & \tabincell{c}{SISO} &Multiple&  \tabincell{c}{FDMA}  & N/A & N/A & \tabincell{c}{\XSolidBrush} &  	\tabincell{c}{\XSolidBrush} & \tabincell{c}{\CheckmarkBold} (FS) &\tabincell{c}{\CheckmarkBold} (FS)&  \tabincell{c}{\XSolidBrush} \\
				
		\hline
		
		[16] & \tabincell{c}{SISO} &Single&\tabincell{c}{N/A}  & Hybrid mode& N/A & N/A & 	\tabincell{c}{\XSolidBrush} & N/A &N/A&  \tabincell{c}{\XSolidBrush} \\

		\hline
		
		[17] & \tabincell{c}{MISO} & Single& \tabincell{c}{N/A}  &  Hybrid mode &  \tabincell{c}{\XSolidBrush} & \tabincell{c}{N/A} &   N/A&  \tabincell{c}{\CheckmarkBold} (SIC) &N/A&  \tabincell{c}{\XSolidBrush} \\
		
		\hline
		
		[18] & \tabincell{c}{SISO} & Multiple& \tabincell{l}{TDMA}  &  Hybrid mode&  N/A & \tabincell{c}{\CheckmarkBold}  &\tabincell{c}{ \CheckmarkBold} & \tabincell{c}{\CheckmarkBold}  (SIC) & 
		\tabincell{c}{ \CheckmarkBold} (TDMA) & \tabincell{c}{\XSolidBrush}  \\
		
		\hline
		
		[19] & \tabincell{c}{SISO} & Single& \tabincell{c}{N/A}  &  Hybrid access &	N/A & N/A  & 
 	\tabincell{c}{\XSolidBrush} &
		\tabincell{c}{ \CheckmarkBold} (SIC)  &  \tabincell{c}{N/A} &  
		\tabincell{c}{\XSolidBrush}    \\

		\hline
		
		[20] & \tabincell{c}{SISO} & Single& \tabincell{c}{N/A}  & Hybrid access & 	N/A & N/A  & 
 	\tabincell{c}{\XSolidBrush} &
		\tabincell{c}{ \CheckmarkBold} (SIC)  &  \tabincell{c}{N/A} &  
		\tabincell{c}{\XSolidBrush}    \\
		
		\hline
		
		[21] & \tabincell{c}{SISO} &Single&  \tabincell{c}{N/A}  & Hybrid access & N/A &	N/A  & 
	\tabincell{c}{\XSolidBrush} &
		\tabincell{c}{ \CheckmarkBold} (MMSE)  &  \tabincell{c}{N/A} &  
		\tabincell{c}{\XSolidBrush}    \\
		
		\hline

		[30] & \tabincell{c}{SISO} &  Multiple & \tabincell{c}{RCMA}  & Hybrid access& N/A &	\tabincell{c}{ \CheckmarkBold} & 
\tabincell{c}{ \CheckmarkBold} &
		\tabincell{c}{ \CheckmarkBold} (SIC)  &  	\tabincell{c}{\XSolidBrush}&  
		\tabincell{c}{ \CheckmarkBold}   \\
		
		\hline
		
		\tabincell{c}{Our\\ work} & \tabincell{c}{SIMO} & Multiple&  \tabincell{l}{SA}  &	Hybrid access & \tabincell{c}{\CheckmarkBold}  & \tabincell{c}{\CheckmarkBold}  & \tabincell{c}{ \CheckmarkBold} & \tabincell{c}{\CheckmarkBold} (SIC)&  \tabincell{c}{\CheckmarkBold} (SA)& \tabincell{c}{\CheckmarkBold} \\
		
		\hline
	\end{tabular}

\end{table*}

In the AmBC/BackCom system, one must consider the double-channel fading experienced by the backscattering signal,  as it contributes to the vulnerability of the signal at the receiver. To elaborate,  both the signals from other backscatter devices (BDs), referred to as backscattering links,  and from the RF source, referred to as direct link (DL),  interfere with the current backscattering signal, leading to significant degradation in signal detection and transmission performance. Therefore, the importance of reducing interference and improving transmission performance in the AmBC/BackCom system cannot be overstated.  In particular, to circumvent the destruction on backscattering transmission imposed by the DL interference (DLI), interference subtraction \cite{Daesena} and opportunistic successive interference cancellation (SIC) \cite{LiZhangFan} have been applied to the AmBC/BackCom as pivotal techniques to deal with the DLI issue. Moreover, in multi-BD scenarios aiming for an interference-free environment, AmBC/BackCom systems often utilize time division multiple access (TDMA), which prevents interference from concurrent transmissions by allowing each BD to backscatter information sequentially \cite{Gu2022, XuGuLi, ZhangLIGaoHan}. Besides, the frequency shift (FS) has also been widely adopted to mitigate the mutual interference (MI) among BDs and the DLI, in which the DL signal and backscattering signals can be shifted to the non-overlapping frequency bands for interference avoidance  \cite{LiLiang2019,Liwcl,ZhangPsigcom}.  
	 
	 
	 Another dark cloud hanging overhead is conventional AmBC/BackCom systems rely heavily on dedicated or surrounding energy sources for backscattering signals of BDs. However, such dedicated energy sources may not always be available, and the surrounding energy signals may not always be stable. One possible solution to circumvent these problems is to allow passive BDs to coexist with traditional active users/systems to share the same spectrum. There have already been some works regarding hybrid active and passive nodes in the spectrum-sharing scenarios for cooperative transmission. In \cite{Li2020TCOM1,Li2020TCOM2,Ye2021}, the BD was incorporated into traditional active transmitter devices so that the adaptive switch between the active and passive mode selection is made possible. By integrating both active and passive devices for spectrum sharing, system performance can be maintained without the need for additional transmit power or channel bandwidth. In \cite{KangLiangYang, Li2020TVT1,Li2020TVT2},  the spectrum sharing between active and passive devices was considered, where the active devices (ADs) served as energy sources for BDs, and BDs were responsible for protecting the ADs from harmful interference similar to the principle of cognitive radio.

	\subsection{Motivations and Contributions}	
	
As observed from the aforementioned works that the traditional centralized access schemes, i.e., TDMA and frequency division multiple access (FDMA), are generally adopted in multi-BD AmBC/BackCom systems to avoid the MI among BDs. However, in view of the proliferating number of IoT devices, is it cost-effective enough to maintain these centralized access schemes? On the one hand, these centralized access schemes operate through an organized and predetermined approach, necessitating coordination among sources. On the other hand, such proactive coordination might not be feasible for simple passive devices. Although a central controller could potentially mitigate this issue, this solution is contingent on two prerequisites. Firstly, the BS must have perfect knowledge of all channel information, without any overhead. Secondly, only a small number of devices should be involved in the optimization. While this approach is well-intentioned, it risks being both ineffective and overly complicated. Thus, it is necessary to explore an access protocol for BackComs with a low overhead and a low complexity.
	
Bearing this in mind, random access has recently been revisited and regarded as a vital technology for the medium access control (MAC) layer of massive IoT \cite{YUj}. Random access presents several advantages: it ensures low latency for small payloads, eliminates the need for initial connection setup, no dedicated resources for connection maintenance, and reduced signaling overhead for resource allocation. Among them, the slotted ALOHA (SA) is indisputably one of the finest protocols for random access, known for mitigating collisions from partially overlapping transmissions while maintaining low complexity. It is noteworthy that the resurgence of SA has been a promising trend, gaining traction across diverse applications, spanning a range of applications such as non-orthogonal multiple access-based systems \cite{HuangSDhieh}, unmanned aerial vehicle-based systems \cite{HadziPe}, satellite networks \cite{ZhaoRen},  wireless-powered communication networks \cite{Iqbal,ChoiShin,HadziPejoski,LiChin}, etc.

In light of these observations, we focus on investigating the performance of the AmBC system with hybrid active and passive multiple access, where one AD and multiple BDs share the same radio spectrum for cooperative information transmission to the receiver.  Besides, incorporating the SA into the multi-BD AmBC system is also one of the centerpieces, in which each BD is accessed by obeying an independent probability. To the best of our knowledge, this is the inaugural integration that the SA is exploited in favour of the access control of BDs, belonging to a decentralized and random access scheme,  in contrast to the existing works that relied on centralized access schemes, such as, TDMA (see, e.g., \cite{Gu2022, XuGuLi,ZhangLIGaoHan,Ye2021}) and FDMA (see, e.g., \cite{LiLiang2019,Liwcl,ZhangPsigcom}). By virtue of its peculiarity, the AmBC system can not only eschew the coordination among passive devices, but also effectively avoid the MI. Although the random code-based multiple access (RCMA) technique was proposed in \cite{HanLiangSun}, where each BD pre-encoded its data by multiplying it with a random spreading code, with a similar objective to ours in terms of avoiding coordination among BDs, the issue of concurrent transmission among multiple BDs, which may lead to decreased transmission rate and decoding performance, was not addressed, which is in contrast to our work.
For a clearer comparison, Table I succinctly summarizes the differences between this paper and the mentioned works. Besides, in contrast to a traditional BS, a battery-constrained AD, its limited power supply and computing capability may not be able to handle overly complex transmit beamforming computations and the energy consumption associated with multiple antenna operations.  In order to circumvent these issues,  the transmit antenna selection (TAS) is adopted to reduce both the power budget and computation complexity.

	In a nutshell, the main contributions of this paper are summarized as follows:
	\begin{itemize}
		\item To investigate and evaluate the performance of the  hybrid active and passive multiple access-based AmBC system with the SA, an optimization problem for maximizing the average throughput is formulated via joint TAS,  beamforming design,  RC adjustment, power control, and channel access probability (CAP) determination,  considering both the non-linearity of energy harvesting (EH) and ensuring transmission fairness among BDs. Besides, the quality-of-service (QoS) support of the AD is also accommodated for benign cooperation. However, the resulting optimization problem is non-convex, presenting significant solution challenges.
		
		\item  To surmount the obstacles posed by the formulated problem, we initially transform the objective function from its max-min form to an equivalent linear one using a slack variable, and then decompose the resulting problem into three sub-problems Then, a block coordinate descent (BCD)-based greedy algorithm,  incorporating maximum ratio combining (MRC), semi-definite programming (SDP), and variable substitution, to derive a sub-optimal solution. It is worth pointing out that the closed-form solutions of the receive beamforming of the backscattering signal, the RC, and the transmit power are derived for analytical insights.
		
		\item Simulation results demonstrate that the proposed algorithm has good convergence and outperforms benchmark algorithms in terms of transmission performance and fairness through an assessment of various network parameters, such as the maximum transmit power, the number of receive antennas, the number of BDs, the noise power, and the minimum achievable throughput on the performance of the different algorithms. Additionally, we pinpoint certain conditions where the efficiency of simpler benchmark algorithms closely matches our proposed algorithm, providing a useful shortcut for practical applications.
		
	\end{itemize}

	\subsection{Organization and Notations}
	
	The remainder of the paper is organized as follows: In Section II, we present the system model and formulate an optimization problem. In Section III, a BCD-based algorithm is proposed to solve the formulated problem. Section IV gives the simulation results, and Section V concludes this paper.
	
	 Throughout the paper, scalars, vectors, and matrices are denoted by lowercase, boldface lowercase, and boldface uppercase letters, respectively. $|\cdot|$ and $||\cdot||$ denote the absolute value of a complex scalar and the $l_2$-norm of a vector, respectively.
$(\cdot)^{\text {H}}$,  Rank$(\cdot)$, and Tr$(\cdot)$ represent the Hermitian,  the rank, and the trace of the matrix, respectively.  $\bm{I}_K$ represents the $K\times K$ identity matrix.  $\mathbb{E}[\cdot]$ is the statistical expectation. $\mathcal{CN}(\mu,\sigma^2)$ is the circularly symmetric complex Gaussian (CSCG) distribution with mean $\mu$ and variance $\sigma^2$. $[\cdot]^+=\max(0,\cdot)$ denotes the  non-negative value. 

		\begin{figure}[t]
		\vspace{-5mm}
	\centerline{\includegraphics[width=2.5in]{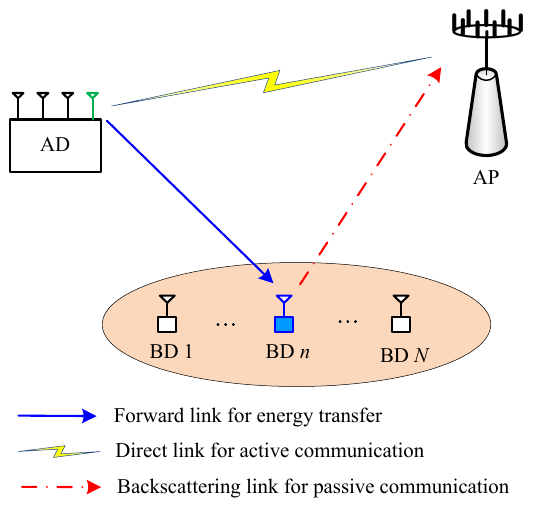}}
	\caption{An uplink AmBC system with multiple BDs.}
	\label{fig1}
\end{figure}

	\section{System Model and Problem Formulation}

	\subsection{Network Architecture and Access Protocol}  
	
   We consider  an  uplink AmBC system, as shown in Fig. \ref{fig1},  which consists of one AD with $M$ transmit antennas (TAs) (i.e., $m\in\mathcal{M}=\{1,2,\cdots,M\}$), $N$ single-antenna BDs (i.e., $n\in\mathcal{N}=\{1,2,\cdots,N\}$), and one access point (AP) with $K$ receive antennas (i.e., $k\in\mathcal{K}=\{1,2,\cdots,K\}$). Specifically, while the AD maintains an active connection with the AP, $N$ BDs cooperate with the AD to passively backscatter information. To reduce the computation complexity and energy consumption, the AD employs a TAS scheme, allowing only one TA to transmit information per slot. Considering that BDs are passive devices and cannot pre-coordinate, we utilize the SA to manage their access.  Assuming that channels between the transmitter and the receiver follow a block-fading model, namely, the corresponding channel gains remain constant for at least one time interval \cite{ChenWang}. Each BD is assumed to always have information to send, in the corresponding transmission duration, the decision of the $n$-th BD on whether to access the channel is determined by the outcome of a Bernoulli experiment, modeled by a random variable $I_n\in \{0,1\}$, which is realized locally at the $n$-th BD \cite{HadziPejoski}. When $I_n=1$ holds, the  $n$-th BD transmits information, and it is silent otherwise, i.e., 
	\begin{equation} \label{a0} 	
		I_n=	\left\{ \begin{aligned}
			& 1, \text {with~probability}~q_n,\\
			& 0, \text {with~probability}~1-q_n,\\
		\end{aligned} \right.
	\end{equation}
	where $\mathbb{E}[I_n]=q_n$, $\mathbb{E}[1-I_n]=1-q_n$, and $q_n$ is the CAP of the $n$-th BD.
	
\begin{figure}[t]
	   \vspace{-5mm}
		\centerline{\includegraphics[width=2.6in]{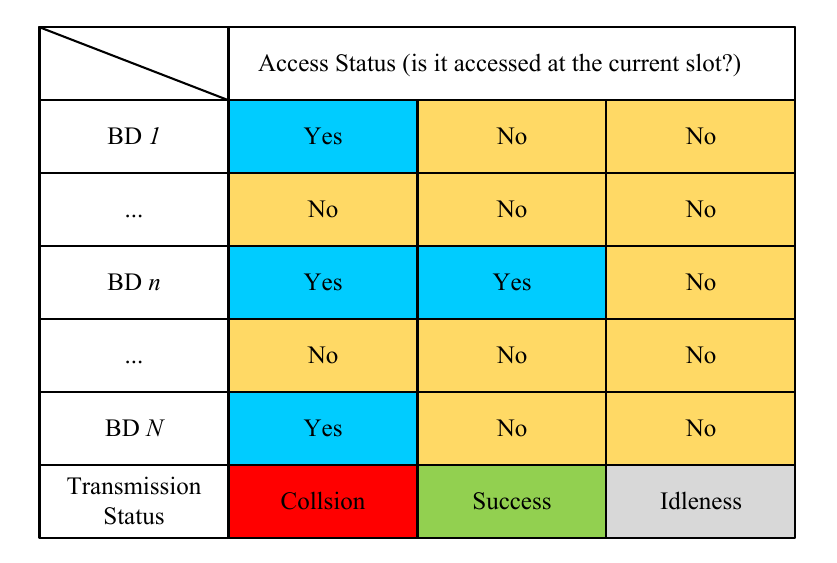}}
		\caption{An access schematic of BDs with the SA.}
		\label{fig1a}
	\end{figure}

	  To prevent MI, only one BD can be accessed simultaneously, as illustrated in Fig. \ref{fig1a}.  Specifically, a transmission collision will occur when two or more BDs transmit information at the same time, and it is regarded as successful transmission if and only if there is only one BD is accessed to backscatter its information. Accordingly, the successful CAP of the $n$-th BD is equal to $\text{Pr}(n)=q_n\prod\limits_{j\neq n}(1-q_j)$.
	
	\subsection{Non-linear Energy Harvesting}
	When the $m$-th TA of the AD is selected, the received signal at the $n$-th BD can be expressed as\footnote{The channel state information (CSI) is can be estimated via training-based techniques \cite{LongLiangGuoYang,ZhangGaoFan}.} 
	\begin{equation} \label{b0}
		y_{m,n}^{\text {BD}}=\sqrt{P_m}h_{m,n}^{\text {f}}s_m+ z_n, \forall m,n, 
	\end{equation}
    where $s_m$ is the symbol transmitted by the $m$-th TA of the AD, satisfying $\mathbb{E}(|s_m|^2)=1$. $P_m$ denotes the transmit power of  the $m$-th TA of the AD.
	$h_{m,n}^{\text {f}}$ represents the channel coefficient from the $m$-th TA of the AD to the $n$-th BD. $z_n\sim\mathcal{CN}(0,\sigma_n^2)$ represents the additive Gaussian white noise (AWGN) with $\sigma_n^2$ indicating the noise power at the $n$-th BD.
	
	It is worth stating that since each BD is passive,  its received signal divides into two components: one for communication requirements and the other as an energy signal for circuit operations. Notably, the noise at each BD, being typically weak for both EH and data backscattering, is frequently overlooked \cite{GuXieLi}.  Thus, based on (\ref{b0}), the harvested power for the operation support of the $n$-th BD from the $m$-th TA can be expressed as
	\begin{equation} \label{s0}
		P_{n}^{\text {EH}}=(1-\alpha_n)\sum\limits_{m=1}^M \beta_mP_m|h_{m,n}^{\text {f}}|^2, \forall n,
	\end{equation}
	where  $\alpha_n$ is the RC of the $n$-th BD.  $\beta_m$ denotes the TAS factor, i.e., when the $m$-th TA is selected, $\beta_m=1$; otherwise, $\beta_m=0$.
	
Extensive studies have demonstrated that wireless EH is a non-linear process that is influenced by the maximum  saturation as well as the minimum sensitivity of the BD's energy harvester (see, i.e., \cite{ WangXiaHuangWu,XuGuGaoLiWu}).
To circumvent resource mismatches,  a non-linear EH model is employed, and thus the harvested energy of the $n$-th BD can be expressed as 
	\begin{equation}  \label{e1}
		\begin{split}
			& \Phi _{n}( P_{n}^{\text{EH}})\\
			&{=}{{\left[\frac{P_{n}^{\text{SA}}}{\exp ({-}{{a}_{n}}P_{n}^{\text{SE}}{+}{{b}_{n}} )}\left\{ \frac{\!1{+}\exp ( {-}{{a}_{n}}P_{n}^{\text{SE}}{+}{{b}_{n}})}{1{+}\exp ( {-}{{a}_{n}}P_{n}^{\text{EH}}{+}{{b}_{n}})}{-}1\right\} \!\right]}^{+}}, \forall n \\ 
		\end{split}
	\end{equation}
	where $\Phi_{n}({{P}_{n}^{\text{EH}}})$  is the actual power that can be harvested by the $n$-th BD. $P_{n}^{\text{SA}}$ and $P_{n}^{\text{SE}}$ are thresholds representing EH saturation and EH sensitivity for the $n$-th BD, respectively. Additionally, the constants $a_n$ and $b_n$ dictate the steepness of $\Phi_{n}({{P}_{n}^{\text{EH}}})$.
	
	\subsection{Hybrid Active and Passive Transmission} 
	
	When the $m$-th TA of the AD is selected  and  the  $n$-th BD is accessed, the  received hybrid active and passive signals at the AP can be expressed as
	\begin{equation} \label{1}
		\bm y_{m,n}^{\text {AP}}{=}\underbrace{
			\sqrt{P_m}\bm h_m^{\text {d}}s_m}_{\text{active link}}{+}\underbrace{\sqrt{\alpha_nP_m}h_{m,n}^{\text {f}}\bm h_n^{\text {b}}s_mc_n}_{\text{passive link}}{+}\bm z_{\text {AP}}, \forall m,n, 
	\end{equation}
     where $c_n$ represents the symbol transmitted by the $n$-th BD with $\mathbb{E}(|c_n|^2)=1$. $\bm h_{m}^{\text {d}}\in \mathbb{C}^{K\times1}$ and $\bm h_n^{\text {b}}\in \mathbb{C}^{K\times1}$ denote the channel vectors of the $m$-th TA-AP link and the $n$-th BD-AP link, respectively. $\bm z_{\text {AP}}\in \mathbb{C}^{K\times1}$ denotes the noise vector of the AP and is assumed to be a  complex Gaussian random vector with $\bm z_{\text {AP}}\sim\mathcal{CN}(0,\sigma_w^2\bm I_K)$.  $\sigma_w^2$ denotes the noise power at the AP.   
	
    Given the double-path loss inherent in the backscattering signal, its signal strength for the backscattering link is significantly weaker compared to the DL. Consequently, the AP decodes the DL signal from $\bm y_{m,n}^{\text {AP}}$ using SIC, cancels the decoded signal, and then detects the backscattering signal $c_n$. 
    Hereinafter, we let $\bm v_{m,n}^{\text {A}} \in  \mathbb{C}^{K\times1}$ and $\bm v_{m,n}^{\text {B}} \in \mathbb{C}^{K\times1}$ denote the receive beamforming employed at the AP for detecting the active signal and the  backscattering signal of the $n$-th BD, respectively, while $\bm v_{m,n}\triangleq[\bm v_{m,n}^{\text {A}}, \bm v_{m,n}^{\text {B}}]$, with adopting normalization \cite{LongLiangPei} such that $||\bm v_{m,n}^{i}||^2=1$ for $i\in \{\text{A},\text{B} \}$. Based on this, the active signal received by the AP can be expressed as
    \begin{equation} \label{2}
    	\begin{aligned}	
    		y_{m,n}^{\text {AP}}=&(\bm v_{m,n}^{\text {A}})^{\text {H}} \bm y_{m,n}^{\text {AP}}\\
    		=&(\bm v_{m,n}^{\text {A}})^{\text {H}} \sqrt{P_m}\bm h_m^{\text {d}}s_m{+}\sqrt{\alpha_nP_m}h_{m,n}^{\text {f}}(\bm v_{m,n}^{\text {A}})^{\text {H}} \bm h_n^{\text {b}}s_mc_n\\
    		+&(\bm v_{m,n}^{\text {A}})^{\text {H}} \bm z_{\text {AP}}, \forall m,n.
    	\end{aligned}
    \end{equation}
    
    Notably, during the decoding of the DL signal, the backscattering signal acts as interference. Thus, the signal-to-interference-plus-noise ratio (SINR) for the AD associated with the $m$-th TA and the $n$-th BD can be represented as 
	\begin{equation} \label{3}
		\begin{aligned}	
			\gamma_{m,n}^{\text {AD}}{=}\frac{|(\bm v_{m,n}^{\text {A}})^{\text {H}}\bm h_m^{\text {d}}|^2P_m}{\alpha_n|h_{m,n}^{\text {f}}|^2|(\bm v_{m,n}^{\text {A}})^{\text {H}}\bm h_n^{\text {b}}|^2P_m{+}||\bm v_{m,n}^{\text {A}}||^2\sigma_w^2}, \forall m,n.
		\end{aligned}
	\end{equation}
	
	 The system transmission time slot is normalized to 1. Accordingly, the achievable throughput of the AD when the $n$-th BD  accesses the system can be expressed as
		\begin{equation} \label{sa1}
		\begin{aligned}	
			R_{n}^{\text {AD}}=\sum\limits_{m=1}^M\beta_m\log_2(1+\gamma_{m,n}^{\text {AD}}), \forall n.
		\end{aligned}
	\end{equation}

	Alternatively, in the scenario where the backscattering signal is being decoded, the DL signal is assumed to be perfectly removed from $\bm y_{m,n}^{\text {AP}}$ using SIC \cite{HanLiangSun,GuLiXie}.  Based on the receive beamforming, the received signal from the $n$-th BD at the AP can be expressed as
	\begin{equation} \label{5}
		\begin{aligned}	
			\bar y_{m,n}^{\text {AP}}=&\sqrt{\alpha_nP_m}h_{m,n}^{\text {f}}(\bm v_{m,n}^{\text {B}})^{\text {H}}\bm h_n^{\text {b}}s_mc_n
			+(\bm v_{m,n}^{\text {B}})^{\text {H}}\bm z_{\text {AP}}, \forall m,n.
		\end{aligned}
	\end{equation}
	
	According to (\ref{5}), the signal-to-noise ratio (SNR) for the $n$-th BD can be expressed as 
	\begin{equation} \label{6}
		\begin{aligned}	
			\gamma_{m,n}^{\text {BD}}=\frac{\alpha_n|h_{m,n}^{\text {f}}|^2|(\bm v_{m,n}^{\text {B}})^{\text {H}}\bm h_n^{\text {b}}|^2P_m}{||\bm v_{m,n}^{\text {B}}||^2\sigma_w^2}, \forall m,n.
		\end{aligned}
	\end{equation}
	
 The average throughput of the $n$-th BD can be obtained as the product of the corresponding achievable throughput and its successful CAP, which yields
	\begin{equation} \label{s2}
		\begin{aligned}	
			R_{n}^{\text {BD}}=\text{Pr}(n)\sum\limits_{m=1}^M \beta_m r_{m,n}^{\text {BD}}, \forall n, 
		\end{aligned}
	\end{equation}
	where  $r_{m,n}^{\text {BD}}=\log_2(1+\gamma_{m,n}^{\text {BD}})$ represents the achievable throughput of the $n$-th BD.

	\subsection{Problem Formulation}

	Our objective is to develop a performance-fair access scheme for multi-BD BackCom systems with the SA  by optimizing the TAS factor and transmit power of the AD, the CAP and RC of each BD, as well as the receive beamforming vectors,  where the QoS requirements of the AD, and the EH demands of BDs are taken into account. Mathematically, the optimization problem can be formulated as	
	\begin{equation} \label{p1} 
		\begin{split}	
			& \underset{\beta_m,  q_n, P_m, \alpha_n, \bm v_{m,n}}{\mathop{\max}}\,  \underset{\forall n}{\mathop{\min}}\, R_{n}^{\text {BD}}\\
			&\text{s.t.}~C_1: \beta_m\in \{0,1\}, \sum\limits_{m=1}^M \beta_m=1,\\
			&\quad ~~C_2: 0<q_n<1, \forall n,\\
			&\quad ~~C_3: R_{n}^{\text {AD}} \ge R_{\min}, \forall n,\\
			&\quad ~~C_4:  \Phi _{n}( P_{n}^{\text{EH}}) \ge P_n^{\text {C}}, \forall n,\\
			&\quad ~~C_5:0<\alpha_n\le1,\forall n, \\ 
			&\quad ~~C_6:\sum\limits_{m=1}^M \beta_mP_m\le P_{\max}, \\ 
			&\quad ~~C_7:||\bm v_{m,n}^i||^2=1, \forall m,n, i,	\\
		\end{split}
	\end{equation} 
	where  $R_{\rm min}$ denotes the minimum required throughput of the AD. $P_n^{\text {C}}$ is the circuit power of the $n$-th BD. $P_{\max}$ represents the maximum transmit power of the AD. It is worth detailing that  $C_1$ constrains the selection factors for TAs, ensuring only one TA is selected for transmission. $C_2$ is used to limit the CAP of each BD.  $C_3$ ensures that the QoS requirements of the AD are met, regardless of which BD is accessed. $C_4$ guarantees that the harvested energy of the $n$-th BD is sufficient to maintain its operation. $C_5$ limits the RC of the $n$-th BD, while $C_6$ constrains the transmit power of the $m$-th TA of the AD. $C_7$ governs the conditions for the receive beamforming vectors.
	
	It should be noted that, we concentrate on exploring the random access for multiple devices in one time slot in this paper, which emphasizes random access competition as opposed to pre-scheduled access as seen in \cite{LiPengHu} and \cite{ZhangGaoFan}, and also contrasts with methods that employ random access across multiple time slots, as discussed in \cite{Iqbal} and \cite{LiChin}. Beyond that, the fairness of the average throughput among different BDs is effectively guaranteed by assigning different CAPs, which can combat the unfairness caused by traditional user selection.
	
	\section{ Algorithm Design}
	
It is clear that problem  (\ref{p1}) is intractable to be solved due to the non-convexity, stemming from the non-smooth objective function and variable coupling. In this section, we devise an effective algorithm to tackle this problem. Specifically,  the max-min form of the objective function is first removed via  a slack variable. Next, we decompose the resulting problem into three sub-problems, addressing each with tailored algorithms. Finally, according to the proposed BCD-based greedy method with MRC and SDP, the sub-optimal solution of problem  (\ref{p1}) is obtained.

	%
	%
	
	\subsection{Problem Reformulation}
The max-min form of the objective function in problem (\ref{p1}) makes it non-smooth and non-convex, posing significant challenges for its solution. To tackle this challenge, we introduce a slack variable $t$ to represent a lower bound on the minimum throughput across all BDs, ensuring $ \underset{\forall n}{\mathop{\min}} R_{n}^{\text {BD}} \ge t$. By doing so, the original multi-objective problem is transformed into a single-objective one, with the goal of maximizing $t$ under the given constraints, which can be equivalently expressed as
	\begin{equation} \label{rp1} 
		\begin{split}	
			& \underset{\beta_m,  q_n, P_m, \alpha_n, \bm v_{m,n}, t}{\mathop{\max}}\, t\\
			&\text{s.t.}~C_1\sim C_7,	\\	
			&\quad ~~C_{8}:  \text{Pr}(n)\sum\limits_{m=1}^M \beta_m r_{m,n}^{\text {BD}}  \ge t, \forall n.	\\	
		\end{split}
	\end{equation} 
	Nonetheless, there are still many obstacles in solving problem (\ref{rp1}), particularly due to the complex interdependencies among optimization variables within the constraints, which involves both binary and continuous variables. Unfortunately,  existing methods cannot be directly applied to yield straightforward solutions. This drives our decision to develop a BCD algorithm as a solution. Specifically, problem (\ref{rp1}) is decomposed into three sub-problems, namely, the sub-problem for beamforming design, the sub-problem for power control and RC optimization, and the sub-problem for CAP determination.
		
	\subsection{The Sub-problem for Beamforming Design}
	With fixed $\beta_m$, $P_m$,  $q_n$, and $\alpha_n$, the sub-problem for beamforming design can be formulated as 
	\begin{equation} \label{p2} 
		\begin{split}	
			& \underset{ \bm v_{m,n}, t}{\mathop{\max}}\,~ t\\
			&\text{s.t.}~C_3, C_7, C_8.	\\	
		\end{split}
	\end{equation} 
Firstly, $\bm v_{m,n}^{\text {B}}$ is considered, which solely involves $C_7$ and $C_8$, and monotonically increases with respect to the objective function. Therefore, to achieve the optimal $\bm v_{m,n}^{\text {B}}$, we can employ the MRC, which can be expressed as
\begin{equation} \label{mrc2} 	
	\begin{aligned}
		(\bm v_{m,n}^{\text {B}})^{*}\triangleq \frac{\bm h_n^{\text{b}}}{||\bm h_n^{\text{b}}||}, \forall n.\\
	\end{aligned}
\end{equation}

Therefore, optimal $t$ for problem (\ref{p2}) can be obtained as 
\begin{equation} \label{mrc3} 	
	\begin{aligned}
	t^*= \underset{\forall n}{\mathop{\min}}\, \text{Pr}(n)\bar r_{m,n}^{\text {BD}},\\
	\end{aligned}
\end{equation}
where $\bar r_{m,n}^{\text {BD}} = \log_2\left (1+\frac{\alpha_n|h_{m,n}^{\text {f}}|^2||\bm h_n^{\text {b}}||^2P_m}{\sigma_w^2}\right)$.

However, $\bm v_{m,n}^{\text {A}}$ still remains unsolved. To deal with it, an optimization problem involved $\bm v_{m,n}^{\text {A}}$ is decomposed, which can be rewritten as 
	\begin{equation} \label{p2a} 
	\begin{split}	
		& \underset{\bm v_{m,n}^{\text {A}}}{\mathop{\max}}\,~ t^*\\
		&\text{s.t.}~C_3, C_7. \\
	\end{split}
\end{equation} 
Unfortunately, problem (\ref{p2a}) remains non-convex owing to the non-linearity of the beamforming vector in $C_3$. To render it solvable, we introduce a new variable, denoted as $\bm V_{m,n}^{\text {A}}=\bm v_{m,n}^{\text {A}}(\bm v_{m,n}^{\text {A}})^{\text {H}}$, allowing problem (\ref{p2a}) to be equivalently rewritten as
	\begin{equation} \label{sp2} \small 
		\begin{aligned}	
			& \underset{ \bm V_{m,n}^{\text {A}}}{\mathop{\max}}\,~ t^*\\
			&\text{s.t.}~C_{3-1}^{\text{BF}}:  \left(2^{R_{\min}}{-}1\right)\left({{\text{Tr} }(\bm H_n^{\text {b}}\bm V_{m,n}^{\text {A}})\alpha_n|h_{m,n}^{\text {f}}|^2P_m{+}\sigma_w^2}\right)\\
			&\quad ~~~~~~~~~\le { {\text{Tr} } (\bm H_m^{\rm d}\bm V_{m,n}^{\text {A}})P_m},\forall n,\\
			&\quad ~~C_{7-1}^{\text{BF}}: {\text{Tr} }(\bm V_{m,n}^{\text {A}})=1,	\forall n, \\		
			&\quad ~~C_{9}^{\text{BF}}:{\rm Rank}(\bm V_{m,n}^{\text {A}})=1, \forall n,	\\	
			&\quad~~C_{10}^{\text{BF}}: \bm V_{m,n}^{\text {A}}\succeq0, \forall n, 
		\end{aligned}
	\end{equation} 
	where $\bm H_m^{\text {d}}=\bm h_m^{\text {d}}(\bm h_m^{\text {d}})^{\text {H}}$ and $\bm H_n^{\text {b}}=\bm h_n^{\text {b}} (\bm h_n^{\text {b}})^{\text {H}}$.
	
Yet despite that, problem (\ref{sp2}) remains difficult to solve due to the non-convex rank-one constraint $C_{9}^{\text{BF}}$. To address this issue, we first omit the constraint $C_{9}^{\text{BF}}$, leading to a standard SDP problem \cite{LuoMa}, i.e.,
		\begin{equation} \label{sp3} 
		\begin{aligned}	
			& \underset{ \bm V_{m,n}^{\text {A}}}{\mathop{\max}}\, ~ t^*\\
			&\text{s.t.}~C_{3-1}^{\text{BF}}, C_{7-1}^{\text{BF}},  C_{10}^{\text{BF}}.
		\end{aligned}
	\end{equation} 
	Note that problem (\ref{sp3}) is  convex and can be solved via CVX \cite{BoydVanden}. The proposed algorithm for solving problem (\ref{sp3}) is summarized in \textbf{Algorithm 1}.
	
		\begin{table*}[t]
		\vspace{-5mm}
		\setcounter{equation}{20}
		\begin{subnumcases} \  
			P_m^*=P_{\max} > \max   \left\lbrace \frac{ \Phi _{n}^{-1}(P_n^{\text {C}})}{|h_{m,n}^{\text {f}}|^2}, \frac {\left(2^{R_{\min}}-1\right) \sigma_w^2}{{\text{Tr} } (\bm H_m^{\text {d}}\bm V_{m,n}^{\text {A}})}\right\rbrace \triangleq P_{m,n}^{\min}, \forall n, \label{21a}\\ 
			\alpha_n^*=\left[ {\min\left\lbrace \underbrace{\frac{{\text{Tr} } (\bm H_m^{\text {d}}\bm V_{m,n}^{\text {A}})P_{\max}-\left(2^{ R_{\min}}-1\right)\sigma_w^2}{(2^{R_{\min}}-1){{\text{Tr} }(\bm H_n^{\text {b}}\bm V_{m,n}^{\text {A}})|h_{m,n}^{\text {f}}|^2P_{\max}}}}_{\alpha_n^{\text {AD}}}, \underbrace{1- \frac{ \Phi _{n}^{-1}(P_n^{\text {C}})}{P_{\max}|h_{m,n}^{\text {f}}|^2}}_{\alpha_n^{\text {EH}}} \right\rbrace}\right]^+, \forall n. \label{21b}
		\end{subnumcases}
		\hrule 
	\end{table*}

\begin{algorithm} [t]
	\caption{The Algorithm for Solving Problem (\ref{sp3})}
	\SetAlgoLined
	\KwIn{Fixed $\alpha_n$, $\beta_m$, $P_m$, $q_n$.}
	\KwOut{$\bm v_{m,n}^{\text {A}}$.}
	
	\KwData {Set the iteration index $i_{\text{A1}}=1$, the tolerance $\omega_{\text{th}}$, the maximum iteration number $ I_{\text{A1}}$.}

	\While{$|\bm V_{ m,n}^{\rm {A}}(i_{\rm{A1}}{+}1){-}\bm V_{m,n}^{\rm {A}}(i_{\rm{A1}})|{\ge} \omega_{\rm{th}}\bm{I}_K{\big|\big|}i_{\rm{A1}}\le I_{\rm{A1}}$}{
		
Solve problem (\ref{sp3}) using  CVX to obtain $\bm V_{m,n}^{\text {A}}(i_{\text{A1}}+1)$.

$i_{\text{A1}}=i_{\text{A1}}+1$.
	}
	
Update $(\bm V_{m,n}^{\text {A}})^*=\bm V_{m,n}^{\text {A}}(i_{\text{A1}})$.

\end{algorithm}

For proceeding, we have the following theorem.
	
	\textbf{\textit{Theorem 1:}} The optimal $(\bm V_{m,n}^{\text {A}})^*$ to problem (\ref{sp3}) satisfies $\text{Rank} ((\bm V_{m,n}^{\text {A}})^*)=1$.
	
	\textbf{\textit{Proof:}}  Please see Appendix A.
	
	\textbf{\textit{Remark 1:}}  According to Theorem 1, the optimal solution $(\bm V_{m,n}^{\text {A}})^*$ for problem (\ref{sp3}) naturally satisfies the rank-one constraint, making it the globally optimal solution for problem (\ref{sp2}). The eigenvalue decomposition of $(\bm V_{m,n}^{\text {A}})^*$ is denoted as $(\bm V_{m,n}^{\text {A}})^*=\bm Q\bm \Lambda \bm Q^{\text {H}}$. Denote the unique non-zero eigenvalue of $(\bm V_{m,n}^{\text {A}})^*$ by $\lambda_{m,n}^*$, and the corresponding eigenvector by $\bm u_{m,n}^*$. The optimal solution to problem (\ref{sp2}) can be obtained as $\sqrt{\lambda_{m,n}^*}\bm u_{m,n}^* $.
	
	\begin{table*}
	\setcounter{equation}{21}
	\begin{equation} \label{p3d} 	
		{\bar t_n}^*=\left\{\begin{aligned}
			&\text{Pr}(n)\log_2\left( 1+{\frac{{\text{Tr} } (\bm H_m^{\text {d}}\bm V_{m,n}^{\text {A}})P_{\max}-\left(2^{ R_{\min}}-1\right)\sigma_w^2}{(2^{R_{\min}}-1){{\text{Tr} }(\bm H_n^{\text {b}}\bm V_{m,n}^{\text {A}})}\sigma_w^2}||\bm h_n^{\text {b}}||^2} \right),  \alpha_n^{\text {AD}}< \alpha_n^{\text {EH}}, \forall n,\\
			&\text{Pr}(n)\log_2\left( 1+\frac{|h_{m,n}^{\text {f}}|^2||\bm h_n^{\text {b}}||^2P_{\max}-||\bm h_n^{\text {b}}||^2{ \Phi _{n}^{-1}(P_n^{\text {C}})}}{\sigma_w^2}\right), \alpha_n^{\text {AD}}\ge \alpha_n^{\text {EH}}, \forall n.\\
		\end{aligned}  \right.
	\end{equation}
	\hrule 
\end{table*}

	\subsection{The Sub-problem for Power Control and RC Optimization}

	With fixed $q_n$, $\beta_m$, and $\bm v_{m,n}$, the sub-problem for power control and RC optimization can be formulated as 
		\setcounter{equation}{19}
	\begin{equation} \label{p3} 
		\begin{split}	
			& \underset{P_m, \alpha_n, \bar t}{\mathop{\max}}\, ~\bar t\\
			&\text{s.t.}~ C_{3-1}^{\text{BF}}, C_5, \\
			&\quad ~~C_{4-1}^{\text{PR}}: (1-\alpha_n)P_m|h_{m,n}^{\text {f}}|^2 \ge \Phi _{n}^{-1}(P_n^{\text {C}}), \forall  n,\\
			&\quad ~~C_{6-1}^{\text{PR}}:P_m\le P_{\max}, \\ 
			&\quad ~~C_{8-1}^{\text{PR}}: \text{Pr}(n)\bar r_{m,n}^{\text {BD}} \ge {\bar t}, 	\forall  n,\\ 
		\end{split}
	\end{equation} 
	where $\bar t$ is a slack variable, $\Phi _{n}^{-1}(x)=[\frac{b_n-\ln A_n}{a_n}]^+$ is the inverse function of $\Phi _{n}$, $A_n=\frac{1{+}\exp ( {-}{{a}_{n}}P_{n}^{\text{SE}}{+}{{b}_{n}})}{B_nx+1}-1$, and $B_n=\frac{\exp ({-}{{a}_{n}}P_{n}^{\text{SE}}{+}{{b}_{n}} )}{P_{n}^{\text{SA}}}$.
	
From an inspection of problem (\ref{p3}), the non-convexity arises due to the tightly coupled variables in constraints $C_{3-1}^{\text{BF}}$, $C_{4-1}^{\text{PR}}$, and $C_{8-1}^{\text{PR}}$. To tackle this, we present the following theorem.
	
	\textbf{\textit{Theorem 2:}} With fixed $q_n$, $\beta_m$, and $\bm v_{m,n}$, the optimal $P_m^*$ and $\alpha_n^*$ can be derived as (\ref{21a}) and (\ref{21b}), respectively.
	
	\textbf{\textit{Proof:}}  Please see Appendix B.

\textbf{\textit{Remark 2:}} The feasible region for $P_{\max}$ is expressed as (\ref{21a}), guiding the setting of the  maximum transmit power of the AD. The value should exceed this lower bound for QoS support.  From (\ref{21b}), it can be extrapolated  that  $\alpha_n^{\text {AD}}$ decreases with the increasing $|h_{m,n}^{\text {f}}|^2$ and/or the decreasing ${\text{Tr} } (\bm H_m^{\text {d}}\bm V_{m,n}^{\text {A}})$. A larger backscattering link or a weaker DL leads to heightened interference for the AD, necessitating reduced RC of the accessed BD to meet the QoS of the AD. Besides, $\alpha_n^{\text {AD}}$  decreases as ${R}_{\min}$ rises. This is because when ${R}_{\min}$ increases, the corresponding SINR for the AD also needs to maintain the same upward trend, which reduces the tolerance for the interference signal (i.e., the backscattering signal). On the other hand, $\alpha_n^{\text {EH}}$ decreases when either $P_{\max}$ or $|h_{m,n}^{\text {f}}|^2$ decrease. This is due to the fact that lower $P_{\max}$ or $|h_{m,n}^{\text {f}}|^2$ reduces the harvested energy for each BD, resulting in an increased ratio of energy allocated to compensate for its circuit. In summary, the scale will be biased towards $\alpha_n^{\text {AD}}$ when the channel gain of  the backscattering link is stronger, or when a higher rate for the AD is desired. Conversely, there is a preference for $\alpha_n^{\text {EH}}$.

	Then, the optimal $\bar t$ for problem (\ref{p3}) can be obtained by substituting (\ref{21a}) and (\ref{21b}) into problem (\ref{p3}), which can be expressed as $\bar t^*=\min{{\bar t_n}^*}$, where ${\bar t_n}^*$ is given by (\ref{p3d}).

	\subsection{The Sub-problem for  CAP Determination} 
	Let us consider $q_n$  in this subsection with fixed $\beta_m$, $\alpha_n$, $P_m$, and $\bm v_{m,n}$. The sub-problem for  CAP determination can be expressed as 
	\setcounter{equation}{22}
	\begin{equation} \label{rp3} 
		\begin{split}	
			& \underset{ q_n, \hat t}{\mathop{\max}}\, ~\hat t\\
			&\text{s.t.}~C_2, 
			C_{8-1}^{\text{CD}}: \text{Pr}(n) \bar r_{m,n}^{\text {BD}}\ge {\hat t}, 	\forall n,\\ 
		\end{split}
	\end{equation}  
where $\hat t$ is a slack variable.

However, multivariate decision is an obstacle to solving problem (\ref{rp3}). To make it treatable,  it can be equivalently transformed into
	\begin{equation} \label{rp3a} 
		\begin{split}	
			& \underset{q_n, \hat t}{\mathop{\max}}\, ~\hat t\\
			&\text{s.t.}~C_2, \\
			&\quad ~~ C_{8-2}^{\text{CD}}:  \ln q_n{+}\sum\limits_{j=1, j\neq n}^N \ln (1-q_j){+}\ln  \bar r_{m,n}^{\text {BD}}\ge {\ln \hat t}, 	\forall n.\\ 
		\end{split}
	\end{equation} 
Regrettably, problem (\ref{rp3a}) is still challenging due to the trickiness of $C_{8-2}^{\text{CD}}$. To deal with it,  we make the following transformation, i.e., 
	\begin{equation} \label{rp3b} 
		\begin{split}	
			& \underset{\bar q_n, \hat q_n, \tilde t}{\mathop{\max}}\, ~\tilde t\\
			&\text{s.t.}~ C_{2-1}^{\text{CD}}: \bar q_n<0, \hat q_n<0, \forall n,  \\
			&\quad ~~ C_{8-3}^{\text{CD}}:\bar q_n{+}\sum\limits_{j=1, j\neq n}^N \hat q_j  \ge \tilde t-\ln \bar r_{m,n}^{\text {BD}},\forall n, \\
			&\quad ~~ C_{11}^{\text{CD}}: \exp(\bar q_n)+\exp(\hat q_n)= 1, \forall n,
		\end{split}
	\end{equation} 
	where $\bar q_n=\ln q_n$, $\hat q_n=\ln(1-q_n)$, and $\tilde t= \ln \hat t$. 
\begin{algorithm} [t]
	\caption{The Algorithm for Solving Problem (\ref{rp3b})}
	\SetAlgoLined
	\KwIn{Fixed $\beta_m$, $\alpha_n$, $P_m$, $\bm v_{m,n}$.}
	\KwOut{$q_n$.}
	
	\KwData Set the iteration index $i_{\text{A2}}=1$, the tolerance $\phi_{\text{th}}$, the maximum iteration number $I_{\text{A2}}$.

	\While{$|\bar t''(i_{\rm{A2}}+1)-\bar t''(i_{\rm{A2}})|\ge \phi_{\rm{th}}{\big|\big|}i_{\rm{A2}}\le I_{\rm{A2}}$}{
		
		Given/Update $\hat q_j(i_{\text{A2}})$, solve problem (\ref{rp3b}) to obtain  $\bar q_n(i_{\text{A2}})$, $\hat q_n(i_{\text{A2}})$, and $\bar t''(i_{\text{A2}})$. 
		
		$i_{\text{A2}}=i_{\text{A2}}+1$.
	}
	
	Calculate $q_n^*{=}\exp(\bar q_n(i_{\text{A2}}))$ and  $(t'')^*{=}\exp(\bar t''(i_{\text{A2}}))$.
	
\end{algorithm}		

\begin{algorithm} [t]
	\caption{The BCD-Based Greedy Algorithm}
	\SetAlgoLined
		\KwIn{$K$, $M$, $N$, $h_{m,n}^{\text {f}}$, $\bm h_m^{\text {d}}$, $\bm h_n^{\text {b}}$, $\sigma_w^2$, $\sigma_n^2$, $R_{\rm min}$, $P_n^{C}$,
			$P_{\max}$, $P_{n}^{\text{SA}}$, $P_{n}^{\text{SE}}$, $a_n$, $b_n$.}
		\KwOut{$P_m$, $\alpha_n$, $\beta_m$, $q_n$, $\bm v_ {m,n}$.}
		
		\KwData {Set the initial point $t_0=0$,  the iteration index $i_{\text{A3}}=1$, the tolerance $\varepsilon_{\text{th}}$, and the maximum iteration number $I_{\text{A3}}$.}
		
		\For{$m=1:M$}{
			
			Set $\beta_m=1$.\
			
			\Do{$|t(i_{\rm{A3}}){-}\bar t(i_{\rm{A3}})|{\le} \varepsilon_{\rm{th}}\&\& |t(i_{\rm{A3}}){-}\hat t(i_{\rm{A3}})|{\le} \varepsilon_{\rm{th}}{\big|\big|} i_{\rm{A3}}> I_{\rm{A3}}$}{
				
				\textbf{Algorithm 1}: Given $\alpha_n(i_{\text{A3}})$, $P_m(i_{\text{A3}})$, and $q_n(i_{\text{A3}})$, solve problems (\ref{p2}) and (\ref{sp3}) to obtain $\bm v_{m,n}(i_{\text{A3}}{+}1)$ and $t(i_{\text{A3}}{+}1)$
				
				\textbf{Closed form}: Given $\bm v_{m,n}(i_{\text{A3}}{+}1)$ and $q_n(i_{\text{A3}})$, solve  problem (\ref{p3}) to obtain  $P_m(i_{\text{A3}}{+}1)$, $\alpha_n(i_{\text{A3}}{+}1)$  and $\bar t(i_{\text{A3}}{+}1)$. 
				
				\textbf{Algorithm 2}: Given $P_m(i_{\text{A3}}{+}1)$, $\bm v_{m,n}(i_{\text{A3}}{+}1)$ and,  $\alpha_n(i_{\text{A3}}{+}1)$, solve problem (\ref{rp3b}) to  obtain $q_n(i_{\text{A3}}+1)$ and $\hat t(i_{\text{A3}}{+}1)$.
				
				$i_{\text{A3}}=i_{\text{A3}}+1$.
			}
			Update $t_{conv}(m)=t(i_{\text{A3}})$.
			
			\eIf{$t_{conv}(m)-t_0\le 0$}{
				Set $\beta_m=0$.}{
				Update $t_0=t_{conv}(m)$.}
	}
\end{algorithm}

So far, problem (\ref{rp3b}) is a standard convex problem, which can be solved efficiently through  the numerical evaluation \cite{BoydVanden}. The details of this algorithm are given by \textbf{Algorithm 2}.

			\begin{figure}[t]
	\vspace{-0mm}
	\centerline{\includegraphics[width=3.5in]{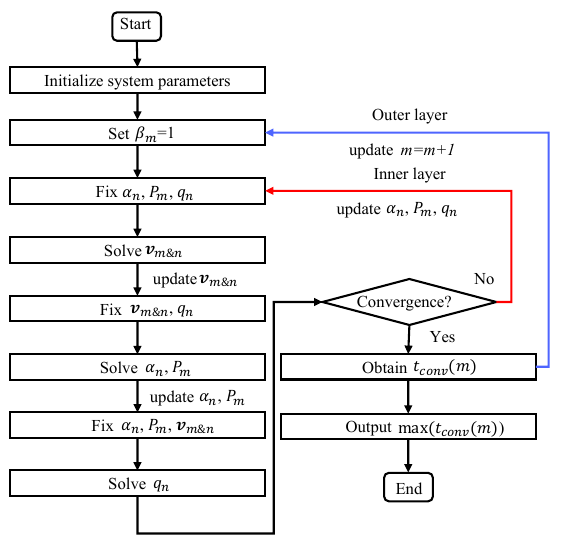}}
	\caption{The flowchart of Algorithm 3.}
	\label{fig1b}
\end{figure}

	\subsection{The Strategy for TAS}
	In the previous subsections, the beamforming vectors, the transmit power of the AD, and the RC and CAP of each BD have been obtained separately. How to select the optimal TA for transmission still remains unsolved. In what follows, we will tackle this issue. The TAS problem is inherently a 0-1 non-linear problem. There have been quite a few existing methods to solve such NP-hard problems, such as greedy algorithm, simulated annealing, penalty dual decomposition method \cite{ShiHong}, and some compressive sensing methods like LASSO \cite{OsbornePresnell} and matching pursuit \cite{BeryhiVagollari}. In this paper, the greedy algorithm is applied to obtain the optimal TAS factor, which is detailed in \textbf{Algorithm 3}, where $t_{conv}(m)$ denotes the convergence point of proposed algorithm under the $m$-th TA.  {Specifically, this algorithm adopts an alternating optimization method to iteratively optimize the sub-problems for beamforming design, power control and RC optimization, and CAP determination in the inner-layer. The outer-layer then obtains the TAS factor through a search process.} For the sake of intuition, a flowchart of \textbf{Algorithm 3} is further provided, as shown in Fig. \ref{fig1b}.

	\subsection{Convergence and Complexity Analysis}
  This subsection delves into the convergence and complexity analysis of the proposed algorithm.  We commence with the convergence analysis of  \textbf{Algorithm 3}. Since the greedy algorithm does not affect the convergence of \textbf{Algorithm 3}, we mainly focus on the convergence of the inner-layer BCD-based algorithm. For completeness, we provide the following theorem.
	
	\textbf{\textit{Theorem 3:}} The convergence of \textbf{Algorithm 3}  is guaranteed.
	
	\textbf{\textit{Proof:}}  Please see Appendix C.
	
	We now turn our attention to evaluating the computational complexity of \textbf{Algorithm 3}.   For the sub-problem for beamforming design, the complexity includes solving the beamforming vectors of the passive and active signals with MRC and \textbf{Algorithm 1}, respectively. Given the number of BDs, the computation for solving the beamforming vector of the passive signal via MRC demands an operation of $\mathcal{O}(N)$. For \textbf{Algorithm 1}, given the number of constraints, interior point method with a matrix variable $\bm V_{m,n}^{\text {A}}$ of size $K\times K$ will take $\mathcal O(\sqrt{NK} \ln(1/\omega_{\text{th}}))$ iterations, with each iteration requiring $\mathcal{O}(N^3K^6)$ arithmetic operations for the worst case, where $\omega_{\text{th}}$ denotes the  precision of the interior point algorithm \cite{Ye}. Therefore, the computational complexity for the beamforming design is $\mathcal{O}_{\text {BF}}= \mathcal{O}(N^{3.5}K^{6.5}\ln(1/\omega_{\text{th}})+N)$. 
	For the sub-problem for power control and RC optimization, solving the transmit power and RC of each BD only requires $\mathcal{O}(1)$ and $\mathcal{O}(N)$ operations, respectively. Thus, we can obtain the computational complexity for this sub-problem is $\mathcal{O}_{\text {PR}}=\mathcal{O}(N)$, ignoring the low-complexity item.
  For the sub-problem for  CAP determination, the computational complexity is $\mathcal{O}_{\text {CD}}=\mathcal O(N^{3.5} \ln(1/\omega_{\text{th}}))$ for operating  \textbf{Algorithm 2}. With $I_{\text{A3}}$ representing the maximum number of alternating iterations, the computational complexity of the inner layer of  \textbf{Algorithm 3} can be obtained as $\mathcal{O}_{\text {BCD}}=\mathcal{O}(I_{\text{A3}}(\mathcal{O}_{\text {BF}}+\mathcal{O}_{\text  {PR}}+\mathcal{O}_{\text {CD}})$. Furthermore, the computational demand for the outer layer of Algorithm 3 stands at $\mathcal{O}(M)$. Overall, the total computational complexity of \textbf{Algorithm 3} can be calculated as $\mathcal{O}(M\mathcal{O}_{\text {BCD}})$.

\begin{figure}[t]
	\centering
	\subfigure[The convergence of the inner layer of Algorithm 3.]
	{
		\includegraphics[width=3.2in]{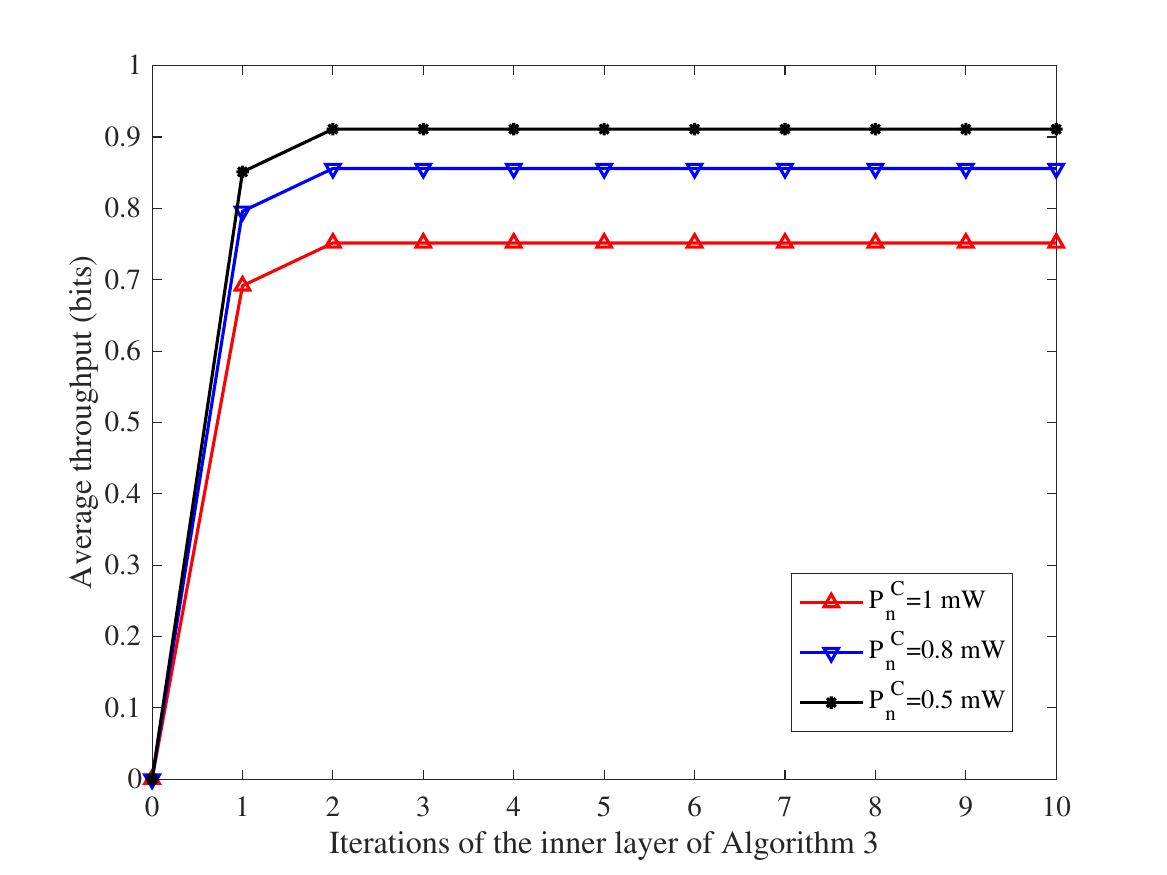} 
	}
	\subfigure[The search process of the outer layer of Algorithm 3.]
	{
		\includegraphics[width=3.2in]{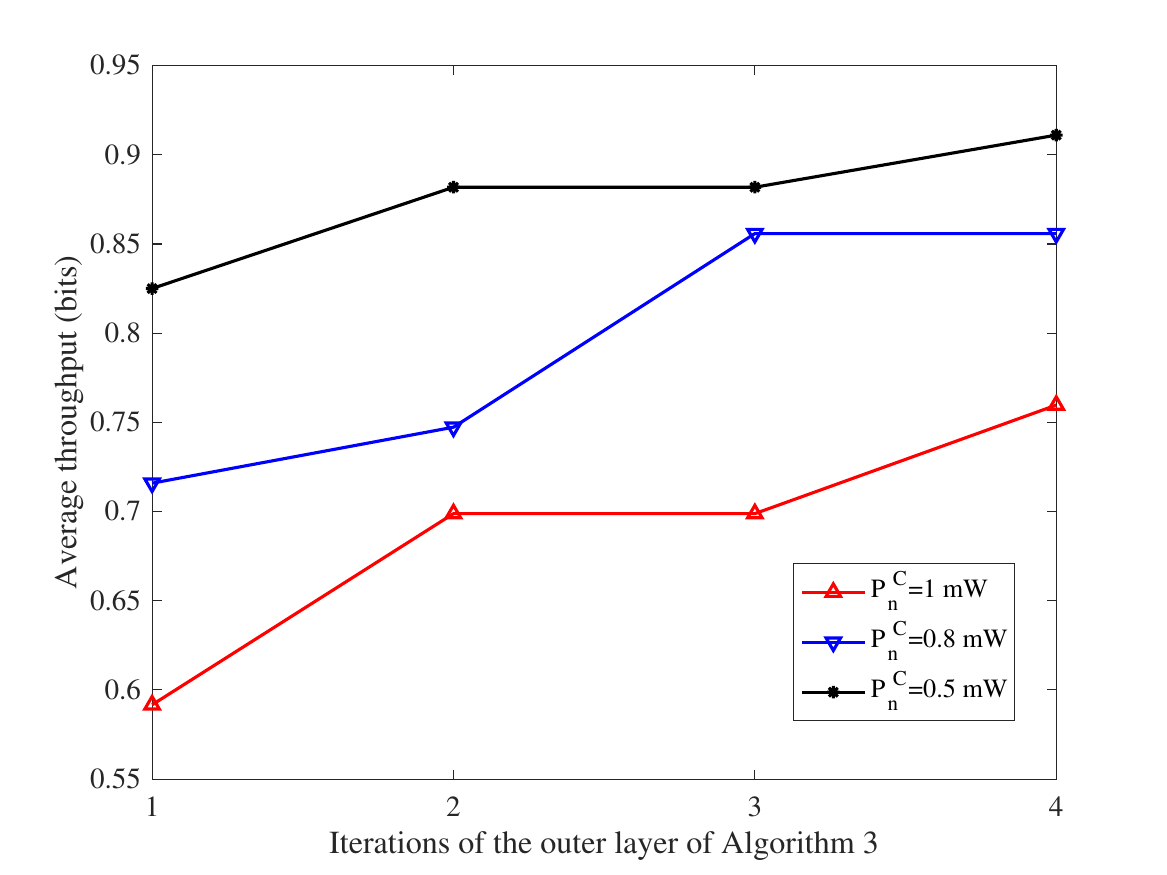} 
	}
	\DeclareGraphicsExtensions.
	\caption{The effectiveness of Algorithm 3.}
	\label{fig2a}
\end{figure}

	\section{Simulation results}
	
	In this section, simulation results are provided to evaluate the performance of the proposed algorithm.  Firstly, we detail the system parameters. Then, we evaluate the proposed algorithm in the following three aspects, namely algorithm effectiveness, algorithm comparison, and algorithm fairness. 
	
	\subsection{Simulation Setup}
	We assume that there are  one AD with 4 transmit antennas, 4 single-antenna BDs, and one AP with 4 antennas in the considered system. The coordinates of the AD and the AP are (0, 0) m and (6, 0) m, respectively. All BDs are randomly distributed within a circle centered at (3, 3) m with a radius of  $r = 2$ m.  Other parameters include $P_{\max}=1$ W, $R_{\min}=1$ bits, $P_n^{\text {C}}=1$ mW, $a_n = 274$, $b_n = 0.29$, $P_n^{\text{SE}}=0.064$ mW,  $P_n^{\text{SA}}=4.927$ mW \cite{WangXiaHuangWu}, and $\sigma_n^2=\sigma_w^2=10^{-8}$ W.	The iteration termination thresholds $\omega_{\text{th}}$, $\phi_{\text{th}}$, and $\varepsilon_{\text{th}}$ are all set as $10^{-3}$.  The maximum iteration numbers $I_{\text{A1}}$, $I_{\text{A2}}$, and $I_{\text{A3}}$ are all set as $10^{3}$.  All simulation results are based on the CVX package \cite{GrantBoyd}. 
	
	Besides, we consider the distance-dependent path loss as large scale fading, and Rician fading as small scale fading for all channels \cite{RamazaniAj}. For example, the channel between the AD and the AP is given by  $\bm h_m^{\text{d}}=\sqrt{\frac{\kappa}{\kappa+1}} \bm h_m^{\text{LoS}}+\sqrt{\frac{1}{\kappa+1}}\bm h_m^{\text{NLoS}}$, where $\kappa=2.8$ is the Rician factor. Here, $\bm h_m^{\text{LoS}} \in \mathbb{C}^{K\times1}$ and $\bm h_m^{\text{NLoS}} \in \mathbb{C}^{K\times1}$ are the line-of-sight (LoS) and non-line-of-sight (NLoS) components of  $\bm h_m^{\text{d}}$, respectively.  Specifically, $\bm h_m^{\text{LoS}}=[1, e^{-j\pi\sin(\theta_i)},\dots,e^{-j\pi(K-1)\sin(\theta_i)}]$, where $\theta_i$ is the the direction of the AD to the AP. Meanwhile,  $\bm h_m^{\text{NLoS}}$ follows the standard Rayleigh fading.  The average power of $\bm h_m^{\text{d}}$ is then normalized by $d_m^{-\mu}$ , where $d_{m}$ is the distance between the AD and the AP, and $\mu= 2.2$ is the path-loss exponent.

%

  	\begin{table}[t]
		\small
	\label{t3}
	\centering
	\caption{$R_{\lambda}$ with different $K$}
		\begin{tabular}{|c|c|c|c|c|c|c|}
			\hline
			BD & 1 &2 & 3 & 4   \\
			\hline
			$K=2$
			&{7.85e-17}  
			& {-2.16e-17}
			& {-3.93e-27}
			& {-1.18e-16}\\
			\hline
			$K=4$
			&{-8.20e{-18}}  
			& {1.47e{-17}}
			& {-3.53e{-17}}
			& {2.02e{-16}}\\
			\hline
	\end{tabular}
\end{table}

	\subsection{The  Algorithm Effectiveness}
In this subsection, we demonstrate the effectiveness of the proposed algorithm by presenting results for various values of $P_n^{\text{C}}$ in Fig. \ref{fig2a}. Fig. \ref{fig2a}(a) demonstrates that the inner layer of the proposed algorithm converges within three iterations, thereby indicating a stable performance. As depicted in Fig. \ref{fig2a}(b), the greedy search method allows for rapid acquisition of the optimal TAS. The optimal antenna selection is indicated by the iteration index where the maximum rate is first achieved. Due to the randomness introduced by the channel fading of the forward link and the DL,  the outcomes of TAS may not always be the same.  Notably, as  $P_n^{\text{C}}$ increases, the average throughput decreases, stemming from the additional harvested energy required to meet elevated circuit power consumption, which in turn reduces the backscattering power.

Additionally, we investigate the rank of the obtained beamforming matrix $(\bm V_{m,n}^{\text {A}})^*$ via the method presented in \cite{SunDerrick}. As depicted in Table III, for different numbers of receive antennas ($K$), the ratio $ R_{\lambda}$ represents the relationship between the second largest eigenvalue and the largest eigenvalue of $(\bm V_{m,n}^{\text {A}})^*$. The results show that the minimum value of $R_{\lambda}$ is consistently close to zero, indicating that the solutions obtained by the proposed algorithm are always rank-one, regardless of the value of $K$. This finding provides further support for the proof presented in Appendix A.

	\subsection{The  Algorithm Comparison}
	
		In this subsection, the comparison of the related algorithms is provided.  To highlight the performance differences, we define and compare the following benchmark algorithms:
	\begin{itemize}
		\item \textit{\textbf{The algorithm with equal CAP (ECAP):}} In this algorithm, each BD is endowed with the same CAP \cite{HuangSDhieh}. This algorithm can be obtained from the proposed algorithm by setting the CAP of each BD as ${1}/{N}$, which can reduce the computation complexity for solving CAP.
		
	\item \textit{\textbf{The algorithm with fixed RC (FRC):}} In this algorithm, the RC of each BD is set to a constant and no longer changes dynamically \cite{RamazaniAj}. This algorithm can be obtained from the proposed algorithm by setting the RC of each BD to a constant.
		
	\item \textit{\textbf{The algorithm with random TAS (RTAS): }}
	    Different from the proposed algorithm, each TA is selected randomly under this algorithm. This algorithm can be obtained from the proposed algorithm by replacing the greedy algorithm with a random approach, which can reduce the computation complexity for solving TAS.
	    
	    \item \textit{\textbf{The algorithm with linear EH (LEH): }}
	    In this algorithm, a linear EH model is adopted \cite{Ye2021}. The saturation and sensitivity of EH are no longer considered, and the process of EH is viewed as a linear one.
	    
        \item \textit{\textbf{The algorithm with TDMA: }} In this algorithm, TDMA is adopted as \cite{XuGuLi}. Here, multiple BDs take turns to backscatter their own signal, leading to unique optimization challenges detailed in Appendix D.
	   
	  \item \textit{\textbf{The algorithm with concurrent transmission (CT): }} In this algorithm, all BDs access the system simultaneously, as described in \cite{HanLiangSun}. This simultaneous access causes significant mutual interference among BDs, differentiating it from the collision-free algorithms mentioned earlier. The unique optimization issues arising from this are detailed in Appendix D.
	    
	\end{itemize}

\subsubsection{\textbf{Comparison for joint optimization}}
	
\begin{figure}[t]
		\centerline{\includegraphics[width=3.5in]{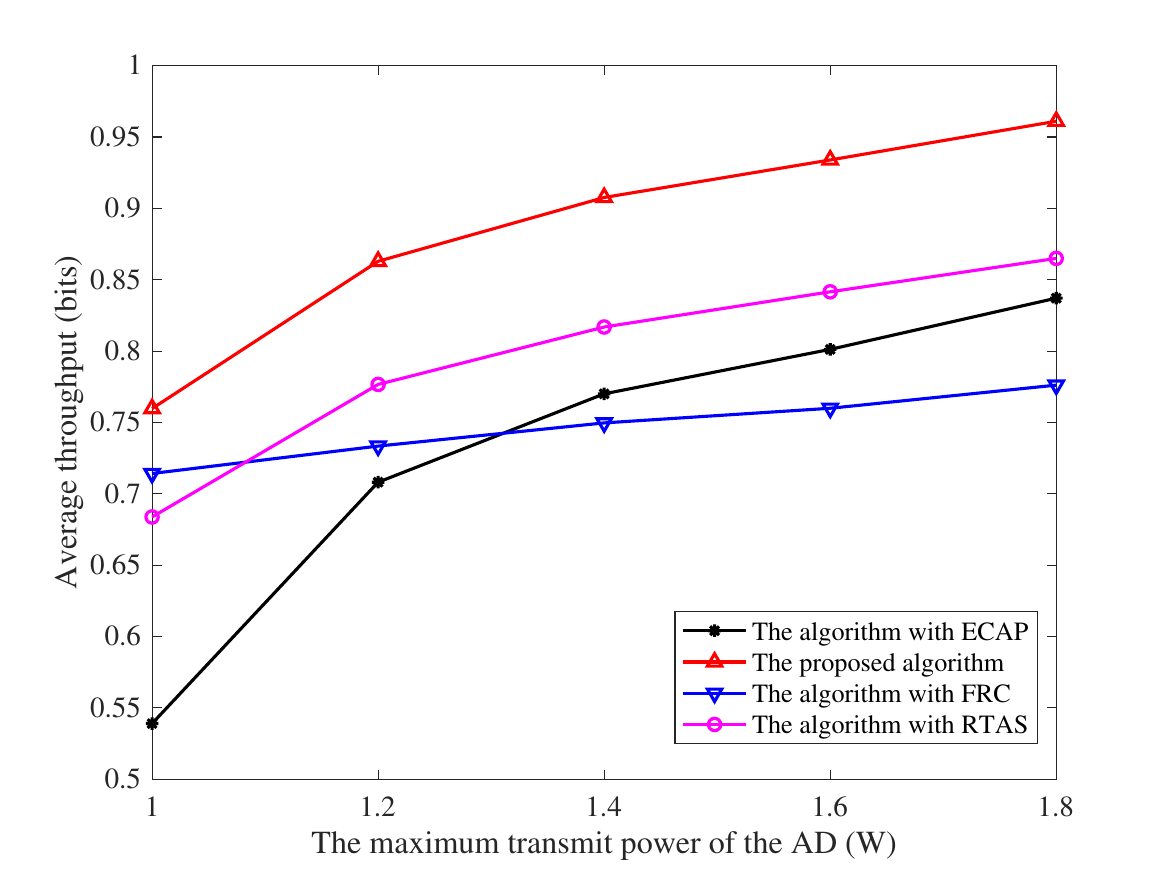}}
		\caption{The average throughput  versus the transmit power of the AD.}
		\label{fig3}
	\end{figure}

		\begin{figure}[t]
	\centerline{\includegraphics[width=3.5in]{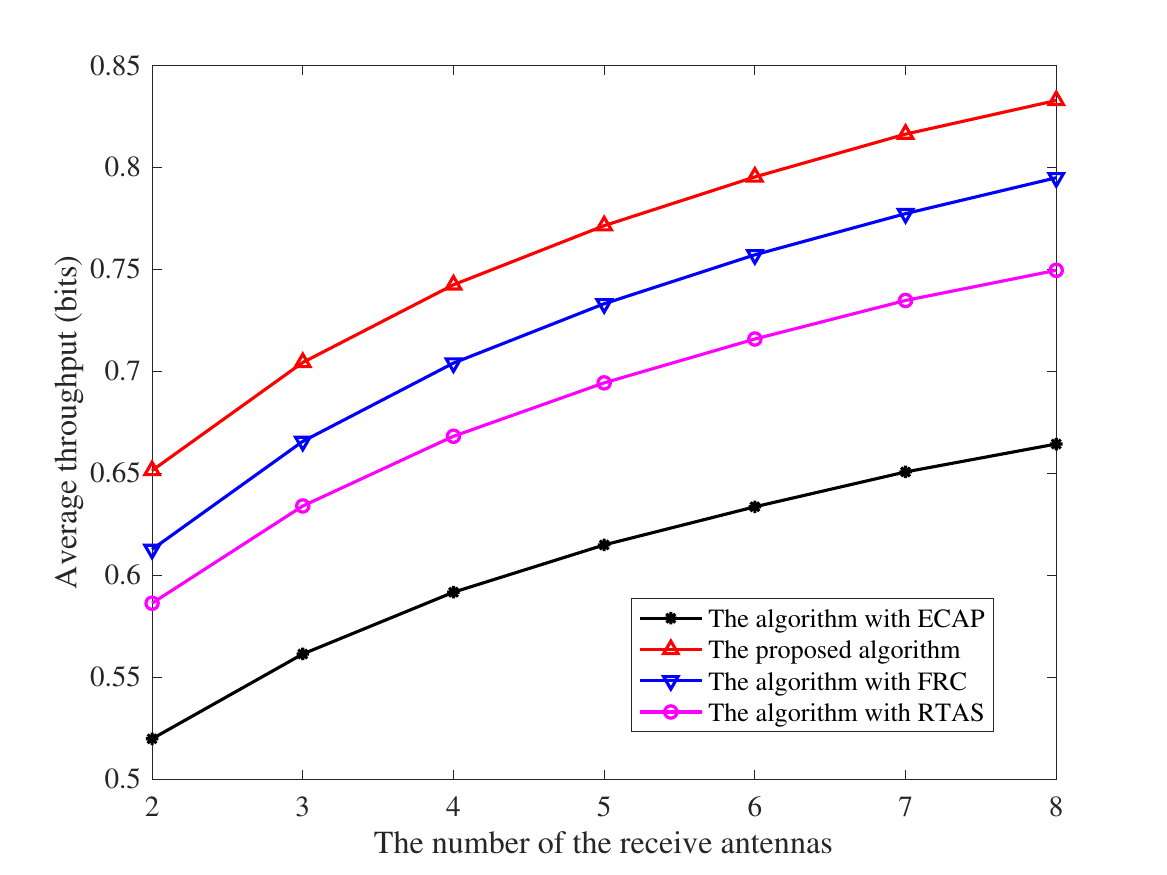}}
	\caption{ The average throughput versus the number of receive antennas.}
	\label{fig4}
\end{figure}

	\begin{figure}[t]
	\centerline{\includegraphics[width=3.5in]{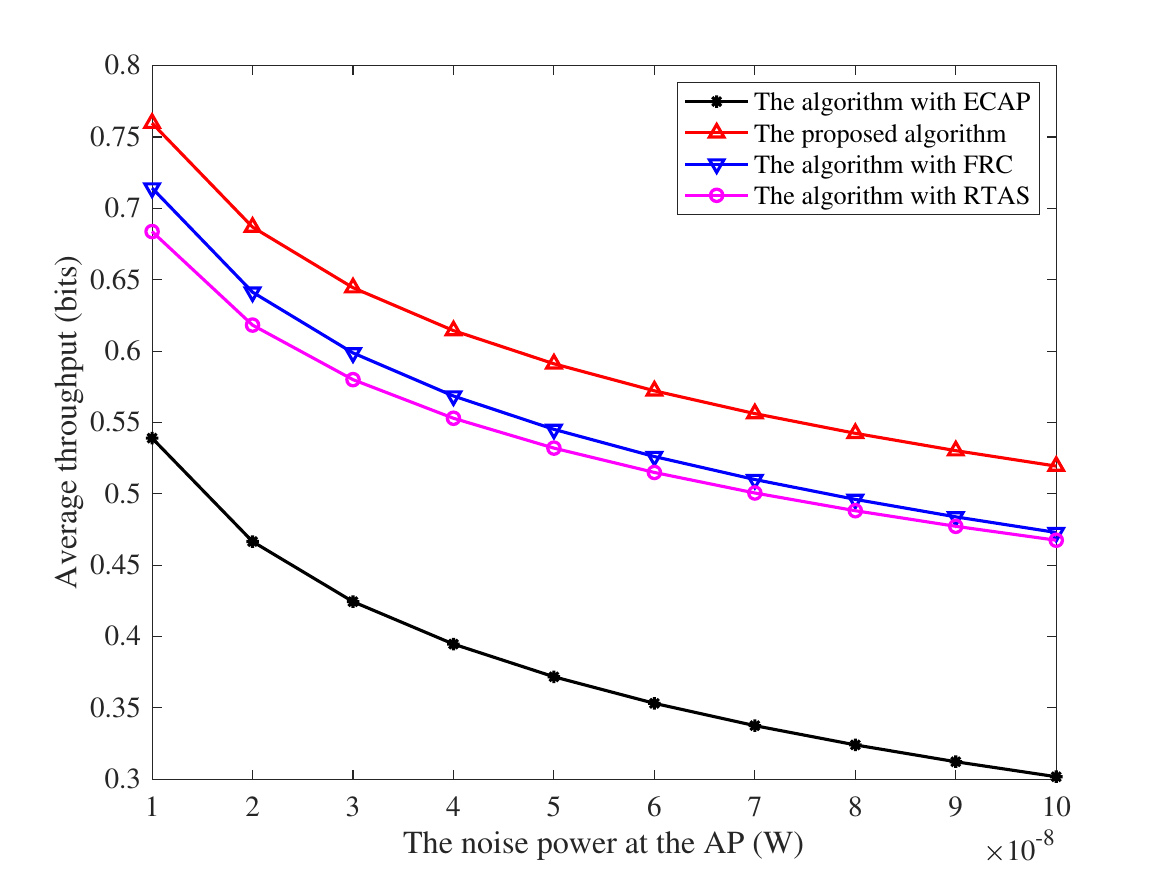}}
	\caption{The average throughput versus the noise power of the AP.}
	\label{fig5}
\end{figure}

	First, we verify the performance of the proposed algorithm by comparing it to benchmarks for evaluating the effectiveness of joint optimization.

	Fig. \ref{fig3} presents the average throughput versus  the maximum transmit power of the AD ($P_{\max}$). As depicted in Fig. \ref{fig3}, the average throughput of all algorithms increases with the rise of $P_{\max}$. This can be attributed to the monotonic increase of the average throughput of the BD (i.e., $R_n^{\text{BD}}$ in (\ref{s2})) with respect to the transmit power. Moreover, the proposed algorithm exhibits a higher average throughput than the other algorithms This is because the algorithms with FRC and ECAP cannot be adjusted dynamically, which results in slight mismatches in resource allocation and is more apparent when $P_{\max}$ is higher. Furthermore, due to its randomized approach, the algorithm with RTAS introduces inherent uncertainty, which may occasionally lead to sub-optimal transmission quality.

Fig. \ref{fig4} reveals the average throughput versus  the number of receive antennas of the AP ($K$). It is observed that the average throughput of all algorithms rises with increasing $K$. This is due to the enhanced beamforming gains with a higher $K$, leading to an increased SNR for the BD. The algorithm with ECAP, however, yields the lowest average throughput. Its emphasis on access fairness (i.e., $q_n$) overlooks the channel differences between BDs. With more receive antennas in play, this oversight becomes even more pronounced, widening the performance gap compared to our proposed algorithm. In contrast to the algorithm with FRC, the proposed method dynamically adjusts BD resource control, preventing energy over-harvesting and thereby boosting the backscattered signal power.

		\begin{figure}[t]
		\centerline{\includegraphics[width=3.5in]{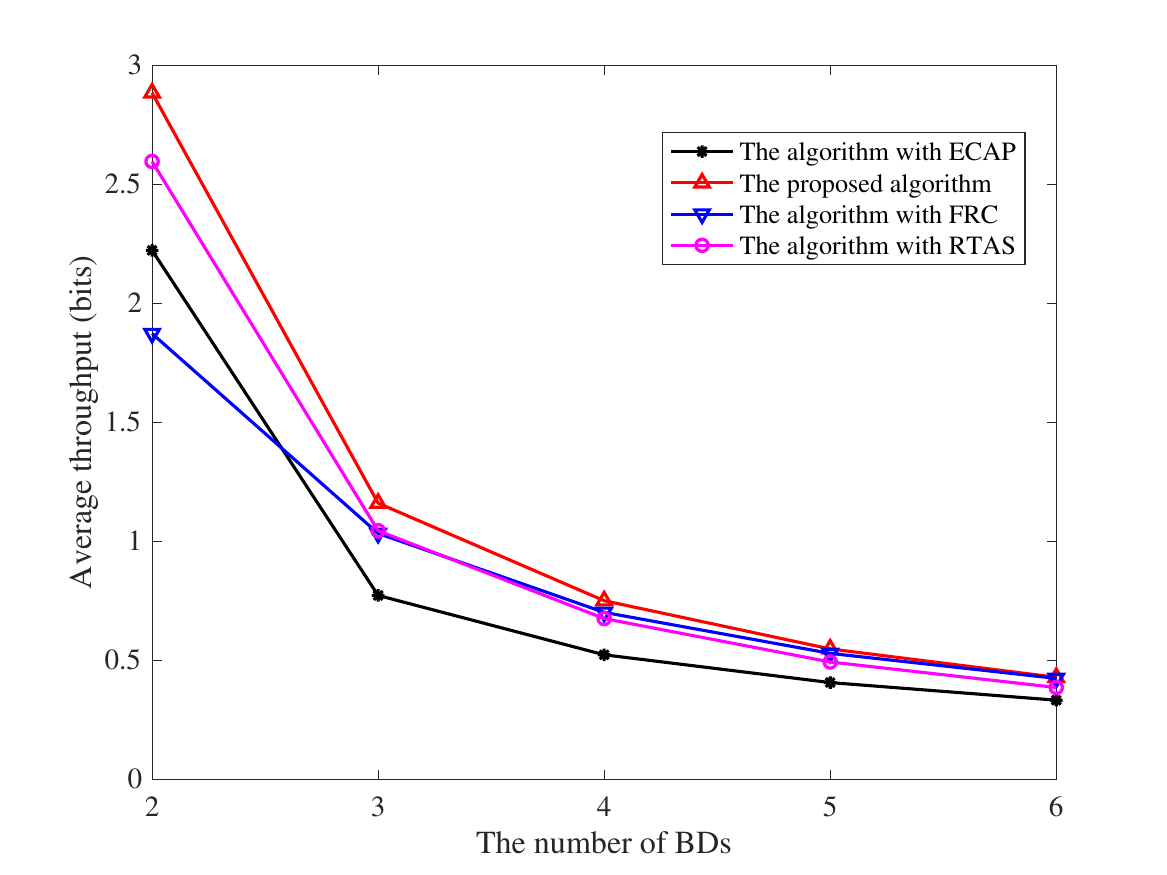}}
		\caption{The average throughput versus the number of BDs.}
		\label{fig6}
	\end{figure}

Fig. \ref{fig5} shows the average throughput versus the noise power at the AP ($\sigma_w^2$). As expected, the average throughput of all algorithms decreases gradually with increasing  $\sigma_w^2$. This is because the higher noise power leads to a decrease in the SNR of both active and passive communications, which is evident from  (\ref{3}) and (\ref{6}), resulting in a reduced achievable throughput. Additionally, as $\sigma_w^2$ increases, the decline in the average throughput of all algorithms becomes more gradual. This is attributed to the non-linear nature of the logarithmic function in $R_n^{\text{BD}}$. Overall, this observation highlights the delicate balance between the background noise and the average throughput.

Fig. \ref{fig6} presents the average throughput versus the number of BDs ($N$). As can be seen from the figure, it is evident that average throughput diminishes as $N$ grows. This can be attributed to the decrease in the CAP of each BD ($q_n$) as $N$ increases, leading to a lower successful CAP (i.e., $\text{Pr}(n)$) and a reduced effective average throughput. Besides, as $N$ gets larger, the gap between different algorithms gradually shrinks. This observation suggests that with a large number of BDs, the performance of the proposed algorithm approaches that of the benchmark algorithms, potentially reducing the computational complexity needed for CAP, RC, and TAS optimization.

\subsubsection{\textbf{Comparison for different mechanism}}

In this part, we are going to evaluate the performance of the proposed algorithm across different transmission methods and EH models. To ensure a fair comparison, we assess the performance of the proposed algorithm against others based on the practical achievable throughput without the CAP form (i.e., $r_{m,n}^{\text{BD}}$) at the AP. Before we discuss performance comparisons, it is important to note that due to the randomness of SA-based algorithms (i.e., the proposed algorithm and the algorithm with LEH), there is variability in the performance displayed among BDs with different CSI. Therefore, to avoid contingency, the results from multiple trials are averaged for these SA-based algorithms.

\begin{figure}[t]
	\centerline{\includegraphics[width=3.5in]{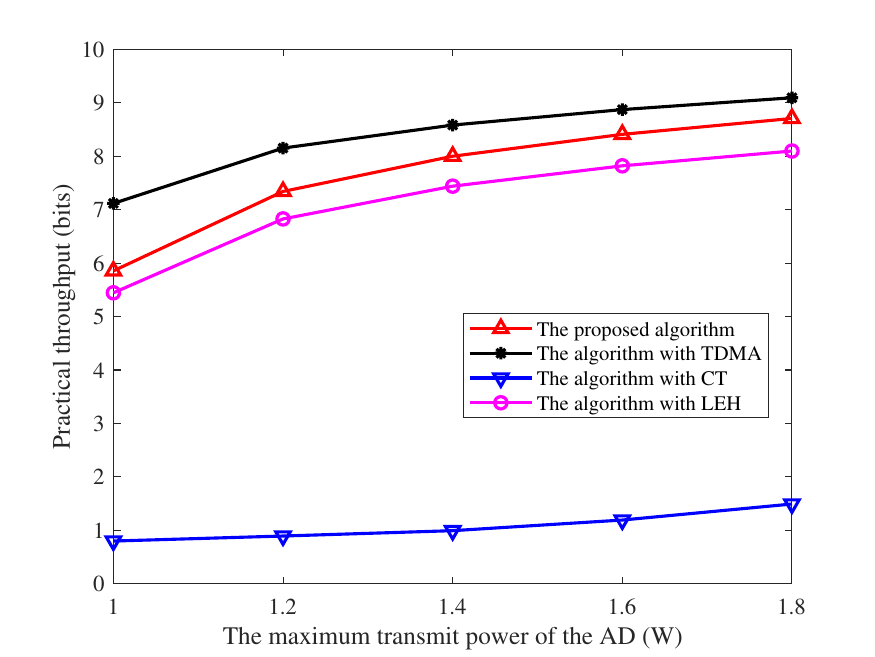}}
	\caption{The practical  throughput  versus the transmit power of the AD.}
	\label{fig3a}
\end{figure}

Fig. \ref{fig3a} demonstrates the practical achievable throughput  versus the transmit power of the AD ($P_{\max}$). It can be observed that the algorithm with CT performs significantly worse than the others. This is due to strong MI caused by the parallel transmission of multiple BDs, which lowers the SINR of each BD, resulting in a poor performance. Conversely, the algorithm with TDMA delivers the best results, with performance increasing as $P_{\max}$ rises, fully exploiting the advantages of centralized management. Additionally, the algorithm with LEH performs lower than the proposed algorithm because it does not account for the non-linear EH effect, leading to the mismatched resource scheduling. Moreover, the proposed algorithm performs near but slightly below the algorithm with TDMA. This discrepancy is due to the prioritization of fairness: BDs with lower achievable throughput are assigned higher CAP, as shown in Table III, reducing the overall average throughput.

\begin{table}[t]
	\label{t4}
	\centering
	\caption{The CAP for 4 BDs \\(*BDs are sorted by achievable throughput in descending order)}
	\begin{tabular}{|c|c|c|c|c|c|c|}
		\hline
		BD & 1st &2nd & 3rd & 4th   \\
		\hline
		CAP
		& 0.18  
		& 0.21
		& 0.40
		& 0.48 \\
		\hline
	\end{tabular}
\end{table}

\begin{figure}[t]
	\centerline{\includegraphics[width=3.5in]{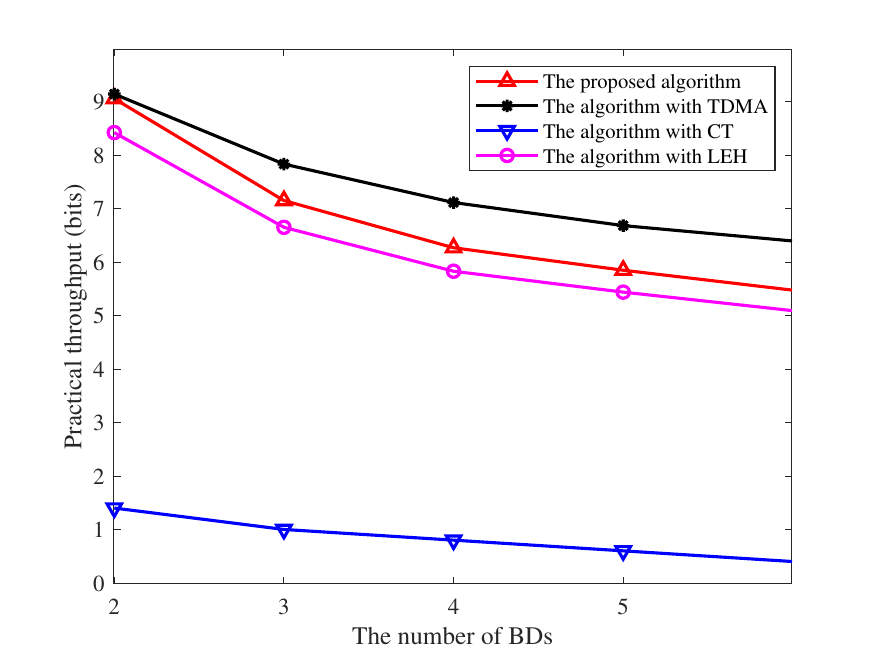}}
	\caption{The practical  throughput versus the number of BDs.}
	\label{fig6a}
\end{figure}

Fig. \ref{fig6a} presents the practical achievable throughput  versus the number of BDs ($N$). As illustrated in the figure, the achievable throughput of all algorithms decreases progressively as $N$ increases. For the algorithm with CT, the increasing $N$ exacerbates MI between BDs, resulting in a low performance. For the algorithm with TDMA, the transmission time for each BD is reduced to meet their QoS requirements, leading to a decline in the system performance. For the SA-based algorithms, the increased access of BDs results in the repeated compression of the CAP for each BD, lowering the overall average performance. Furthermore, the algorithm with TDMA outperforms the others because it avoids interference while reasonably controlling the access of all BDs in a centralized manner. Due to random access, the average value of the proposed algorithm is reduced by the performance differences among different access BDs.

Overall, by using a collision-free transmission scheme, the algorithm with TDMA and the SA-based algorithms significantly enhance BackCom system transmission, standing out against the algorithm with TC, highlighting that interference has a greater impact on weak-signal transmission systems such as BackCom. It is worth noting that although TDMA achieves a superior performance, its implementation requires centralized control and accurate time synchronization to manage access switching, which incurs a significant overhead. In contrast, the proposed algorithm relies on random decisions and has a lower overhead compared to TDMA settings.

\subsection{The Algorithm Fairness}

	\begin{figure}[t]
	\centerline{\includegraphics[width=3.5in]{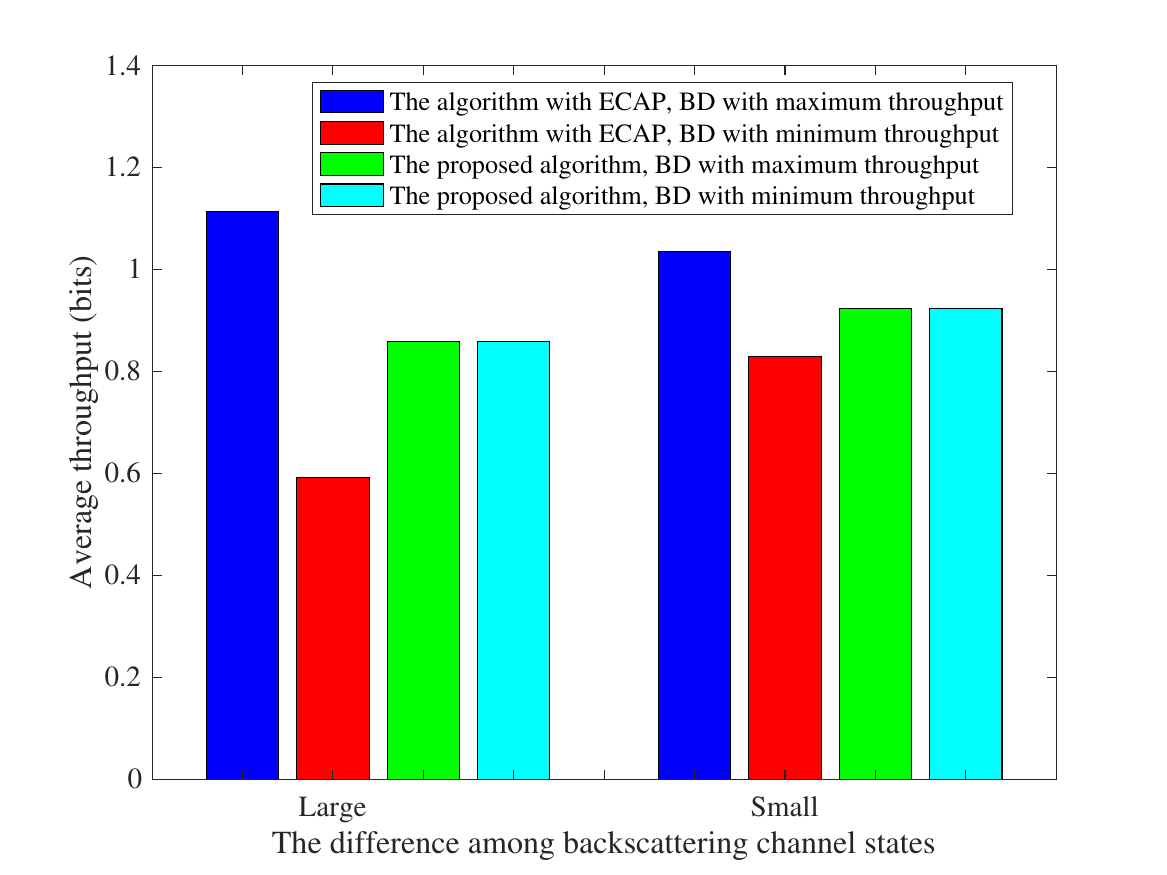}}
	\caption{The average throughput  versus the different backscattering channel state.}
	\label{fig8}
\end{figure}

\begin{figure}[t]
	\centerline{\includegraphics[width=3.5in]{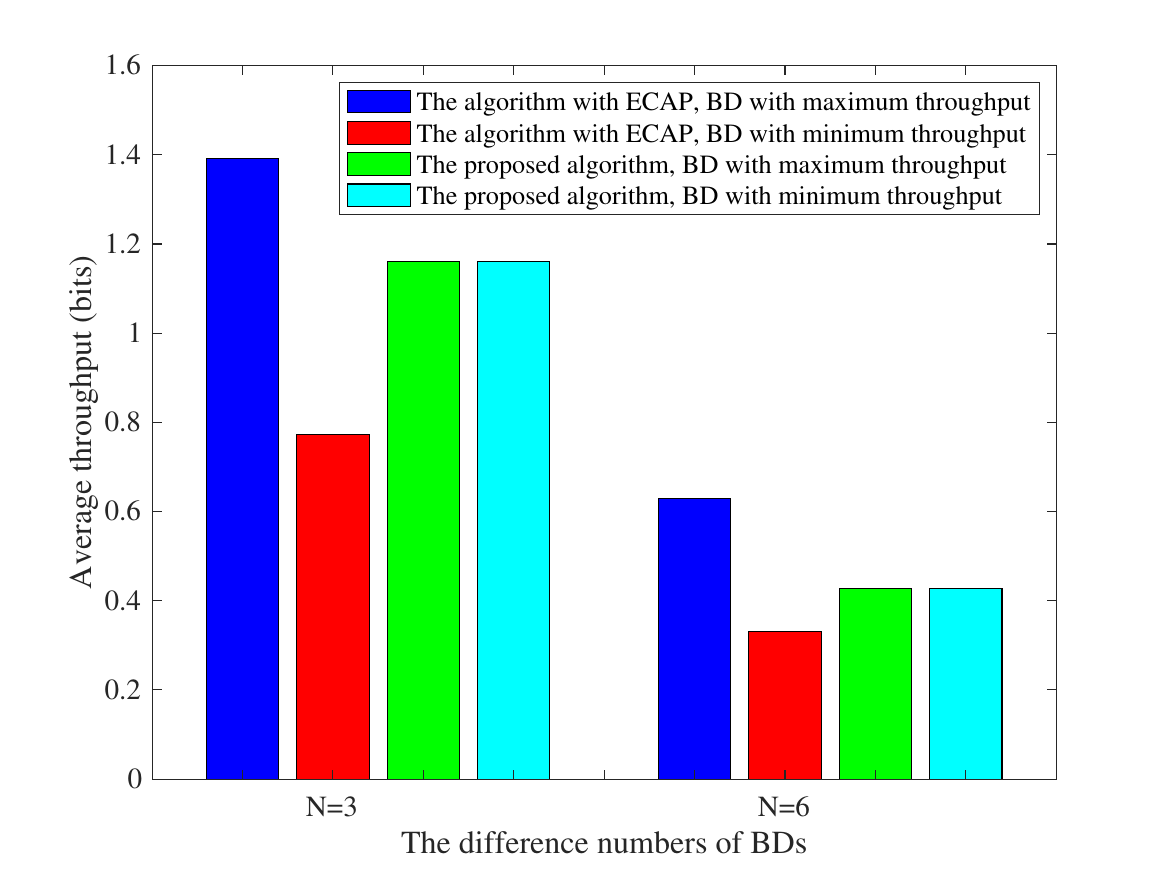}}
	\caption{The average throughput  versus the different number of BDs.}
	\label{fig9}
\end{figure}
	
This subsection contrasts the performance of the proposed algorithm with the ECAP algorithm, evaluating fairness through backscattering channel states and the number of BDs.
	
Fig. \ref{fig8} illustrates the average throughput against varying backscattering channel states ($h_{m,n}^{\text {f}} \bm h_n^{\text {b}}$), as described in \cite{XuXieWu}.  For smaller differences in $h_{m,n}^{\text {f}} \bm h_n^{\text {b}}$ across BDs, both algorithms exhibit a minimal throughput variance; in fact, the variance for the proposed algorithm nears zero. However, as the differences widen, the throughput variance for ECAP increases, whereas the proposed algorithm maintains a tighter spread.  This is because the algorithm with ECAP overlooks channel discrepancies among users, causing resource mismatches. In contrast, the proposed algorithm prioritizes balanced performance, marking a key distinction.

Fig. \ref{fig9} shows the average throughput versus the number of BDs ($N$). As depicted, with an increasing $N$, the average throughput declines, a trend also evident in Fig. \ref{fig6}. Notably, the proposed algorithm maintains nearly equal throughput across all BDs, irrespective of $N$. In contrast, the  algorithm with ECAP reveals pronounced throughput disparities among BDs. This distinction arises because our proposed algorithm employs a max-min form, ensuring optimal fairness in resource distribution.
	
Comparing both figures, the proposed algorithm clearly promotes superior transmission fairness, especially when ignoring channel disparities and BD numbers, a stark contrast to ECAP. However, this commitment to fairness introduces a trade-off: while fairness is prioritized, the peak throughput of the system  may be compromised, as seen when contrasting the maximum throughput of the proposed algorithm  with  that  of  the algorithm with ECAP.
	
	\begin{figure}[t]
		\centerline{\includegraphics[width=3.5in]{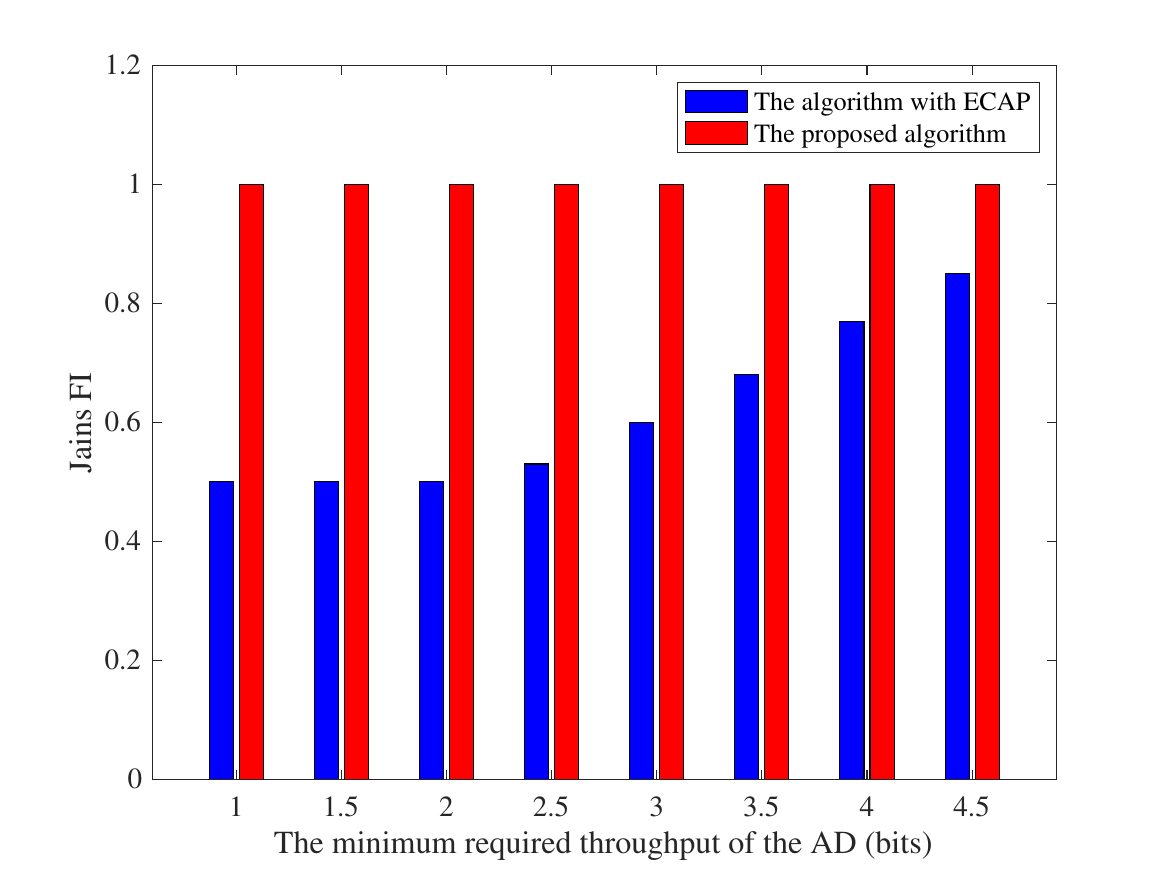}}
		\caption{The Jain's FI  versus the minimum throughput of the AD.}
		\label{fig7}
	\end{figure}

  Furthermore, in order to assess the fairness of the two algorithms more accurately, we employ Jain’s fairness index (FI) as given by ${\text{FI}}=(\sum\limits_{n=1}^N R_n^{\text{BD}})^2/(N\sum\limits_{n=1}^N(R_n^{\text{BD}})^2)$, as referenced in \cite{XuGuLiYang}. 
Specifically, Fig. \ref{fig7} showcases how Jain's FI responds to varying levels of minimum required throughput of the AD ($R_{\min}$). As we observe an increase in $R_{\min}$, the FI for the ECAP algorithm initially remains stable, then starts to climb. This behavior is attributed to the corresponding rise in the AD's SINR as $R_{\min}$ grows, which in turn suppresses the backscattered signal from the accessed BD. One immediate consequence is the constraint imposed on RC by $\alpha_n^{\text {AD}}$, as shown in (\ref{21b}).  Consequently, the throughput disparity between BDs narrows, enhancing the fairness of this algorithm. It is worth noting that the FI of our proposed algorithm consistently hovers near 1, signaling optimal fairness. Drawing insights from the figure, when a higher $R_{\min}$ is necessary, the algorithm with ECAP could serve as an approximation to our proposed method, affording a reduction in computational complexity, especially in the optimization of CAP represented as $q_n$.

	\section{Conclusions}
	This paper investigates and analyzes the hybrid active-passive multiple-access  BackCom system, where the SA-based protocol is introduced to overcome the high overhead issue of traditional multi-access protocols in dealing with massive devices, and an optimization problem for throughput fairness is formulated. We then take the objective function, structured in a max-min form, and transform it into an equivalent linear one using a slack variable. Subsequently, this reformulated problem is decomposed into three distinct sub-problems. Building on this, we propose a BCD-based algorithm that integrates MRC, SDP, and variable substitution techniques to obtain the sub-optimal solution.   Simulation results unequivocally show that the proposed algorithm outperforms benchmark ones, both in terms of transmission performance and fairness.

	\begin{appendices}
		
	\section{Proof of  Theorem 1}

				\begin{table*}
	\vspace{2mm}
	\begin{equation} \label{apa1} 	
		\begin{aligned}
			&\mathcal{L}(t^*, \bm V_{m,n}^{\text {A}}, \chi_{m,n}, \varepsilon_{m,n}, \bm \Omega_{m,n})=t^*+\sum\limits_{n=1}^N \chi_{m,n}  \left\lbrace { {\text{Tr} } (\bm H_m^{\text {d}}\bm V_{m,n}^{\text {A}})P_m-\left(2^{R_{\min}}{-}1\right)({{\text{Tr} }(\bm H_n^{\text {b}}\bm V_{m,n}^{\text {A}})\alpha_n|h_{m,n}^{\text {f}}|^2P_m{+}\sigma_w^2})}\right\rbrace\\
			&+\sum\limits_{n=1}^N \varepsilon_{m,n}(1-{\text{Tr} }(\bm V_{m,n}^{\text {A}})) +\sum\limits_{n=1}^N {\text{Tr} } (\bm V_{m,n}^{\text {A}}\bm \Omega _{m,n})=t^*+\sum\limits_{n=1}^N{\text{Tr} }(\bm V_{m,n}^{\text {A}}	\bm X_{m,n} )-\sum\limits_{n=1}^N\chi_{m,n}\left(2^{ R_{\min}}{-}1\right)\sigma_w^2 +\sum\limits_{n=1}^N\varepsilon_{m,n}.\\
		\end{aligned}
	\end{equation}
	\hrule 	
\end{table*}

			\begin{table*}
	\setcounter{equation}{30}
	\begin{equation} \label{ap1} 	
		\begin{aligned}
			&P_m\ge \max   \left\lbrace \frac{ \Phi _{n}^{-1}(P_n^{\text {C}})}{(1-\alpha_n)|h_{m,n}^{\text {f}}|^2}, \frac {\left(2^{R_{\min}}-1\right) \sigma_w^2}{{\text{Tr} } (\bm H_m^{\text {d}}\bm V_{m,n}^{\text {A}})-\left(2^{R_{\min}}-1\right){\text{Tr} }(\bm H_n^{\text {b}}\bm V_{m,n}^{\text {A}})\alpha_n|h_{m,n}^{\text {f}}|^2}\right\rbrace. 
		\end{aligned}
	\end{equation}
	\hrule 
\end{table*}
		
	The Lagrangian function pertaining to problem (\ref{sp3}) is expressed as (\ref{apa1}), where $\chi_{m,n}$ and $\varepsilon_{m,n}$ denote the dual variables associated with $C_{3-1}^{\text{BF}}$, $C_{7-1}^{\text{BF}}$,  $\bm \Omega_{m,n}$ is the dual matrix associated with the positive semi-definite (PSD) constraint $C_{10}^{\text{BF}}$, and $\bm X_{m,n}$ is expressed as 
			\setcounter{equation}{26}
				\begin{equation} \label{apa1a} 	\small 
		\begin{aligned}
			\bm X_{m,n}=&\chi_{m,n} \left\lbrace \bm H_m^{\text {d}}P_m-\left(2^{R_{\rm min}}{-}1\right)\bm H_n^{\text {b}}\alpha_n|h_{m,n}^{\text {f}}|^2P_m\right\rbrace \\
			&-\varepsilon_{m,n}+\bm \Omega_{m,n}.
		\end{aligned}
	\end{equation}

For the optimal solution to our problem, we can leverage the Karush-Kuhn-Tucker (KKT) conditions. Assuming $(\bm V_{m,n}^{\text {A}})^*$, $\chi_{m,n}^*$, $\varepsilon_{m,n}^*$, and $\bm \Omega_{m,n}^*$ are the optimal original and dual variables respectively, the KKT conditions can be expressed as	
		\begin{equation} \label{apa5} \small 
	\begin{aligned}
		&\bigtriangledown_{\bm V_{m,n}^{\text {A}}}  \mathcal{L}((\bm V_{m,n}^{\text {A}})^*,  \chi_{m,n}^*, \varepsilon_{m,n}^*, \bm \Omega_{m,n}^*)=	\bm X_{m,n}^*=\bm 0,
	\end{aligned}
\end{equation}
and
		\begin{equation} \label{apa5a} 
	\begin{aligned}
	(\bm V_{m,n}^{\text {A}})^*\bm \Omega_{m,n}^*=\bm 0,
	\end{aligned}
\end{equation}
where $\bm X_{m,n}^*=\bm \Omega_{m,n}^*-\varepsilon_{m,n}^*+\chi_{m,n}^* \bm H_m^{\text {d}}P_m-\chi_{m,n}^*\left(2^{R_{\min}}{-}1\right)\bm H_n^{\text {b}}\alpha_n|h_{m,n}^{\text {f}}|^2P_m$.

By multiplying (\ref{apa5}) on both sides by $(\bm V_{m,n}^{\text {A}})^*$,  and substituting (\ref{apa5a}) into the obtained multiplied equation,  we obtain the following equation, i.e., 
		\begin{equation} \label{apa6} 
	\begin{aligned}
 \bm Y_{m,n}^*(\bm V_{m,n}^{\text {A}})^*=\bm 0,
	\end{aligned}
\end{equation}	
where 
		$\bm Y_{m,n}^*=\chi_{m,n}^* \bm H_m^{\text {d}}P_m-\varepsilon_{m,n}^*-\chi_{m,n}^*\left(2^{R_{\min}}{-}1\right)\bm H_n^{\text {b}}\alpha_n|h_{m,n}^{\text {f}}|^2P_m$.

From (\ref{apa6}), we observe that $\text{Rank} (\bm Y_{m,n}^*)+\text{Rank} ((\bm V_{m,n}^{\text {A}})^*)\le K$. Given that the channels are independently distributed, we can verify that $\text{Rank} (\bm Y_{m,n}^*)$ is ($K-1$) at least. This implies that $\text{Rank} ((\bm V_{m,n}^{\text {A}})^*)\le 1$.  Conversely, to meet the QoS requirements, we need $(\bm V_{m,n}^{\text {A}})^*\neq\bm 0$ (or equivalently, $\text{Rank} ((\bm V_{m,n}^{\text {A}})^*)\ge 1$). Therefore, $\text{Rank} ((\bm V_{m,n}^{\text {A}})^*)=1$, completing the proof.

	\section{Proof of  Theorem 2}	
    The objective function of problem (\ref{p3}) can be readily shown to be a monotonically increasing function with respect to $P_m$ and $\alpha_n$.
	Consequently, irrespective of the chosen TA, its optimal transmit power is $P_m^*=P_{\max}$. Besides, according to $C_{3-1}^{\text{BF}}$ and $C_{4-1}^{\text{BF}}$, we have the lower bound for $P_m$, which is expressed as (\ref{ap1}).
	Then, let $\alpha_n=0$, which yields the minimum required transmit power of the AD when the harvested energy of the BD exactly compensate for its circuit power and no information is backscattered. From this, we can derive
			\setcounter{equation}{31}
	\begin{equation} \label{p3ba} 
		\begin{aligned}
			&P_{m,n}^{\min}= \max   \left\lbrace \frac{ \Phi _{n}^{-1}(P_n^{\text {C}})}{|h_{m,n}^{\text {f}}|^2}, \frac {\left(2^{R_{\min}}-1\right) \sigma_w^2}{{\text {Tr}} (\bm H_m^{\text {d}}\bm V_{m,n}^{\text {A}})}\right\rbrace, \forall n.
		\end{aligned}
	\end{equation} 
	
Conversely, based on constraints $C_{3-1}^{\text{BF}}$, $C_{4-1}^{\text{PR}}$, and $C_5$, it can be deduced that
	\begin{equation} \label{p3c} \small 	
		0\le {\alpha_n}\le \left\{ \begin{aligned}
			& \frac{{\text{Tr} } (\bm H_m^{\text {d}}\bm V_{m,n}^{\text {A}})P_{\max}-\left(2^{R_{\min}}-1\right)\sigma_w^2}{\left(2^{R_{\min}}-1\right){{\text {Tr}}(\bm H_n^{\text {b}}\bm V_{m,n}^{\text {A}})|h_{m,n}^{\text {f}}|^2P_{\max}}}, \forall n,\\
			& 1- \frac{ \Phi _{n}^{-1}(P_n^{\text {C}})}{P_{\max}|h_{m,n}^{\text {f}}|^2}, \forall n. \\
		\end{aligned}  \right.
	\end{equation}
	Hence, given its monotonicity, the optimal RC $\alpha_n^*$ can be determined by (\ref{21b}). In accordance with the above conclusions, the proof is complete.
	
		\section{Proof of  Theorem 3}
		
	 Defining $\left\lbrace \alpha_n(i_{\text{A3}}), P_m(i_{\text{A3}}),  q_n(i_{\text{A3}}), \bm v_{m,n}(i_{\text{A3}}) \right\rbrace $ as the $i_{\text{A3}}$-th iteration solution of problems (\ref{p2}), (\ref{sp3}), (\ref{p3}), and (\ref{rp3b}), the objective function of problem (\ref{p1}) can be calculated as $t_{\text{In}}(\alpha_n(i_{\text{A3}}), P_m(i_{\text{A3}}),  q_n(i_{\text{A3}}), \bm v_{m,n}(i_{\text{A3}}))=\underset{\forall n}{\mathop{\min}}\, R_{n}^{\text {BD}}(\alpha_n(i_{\text{A3}}), P_m(i_{\text{A3}}),  q_n(i_{\text{A3}}), \bm v_{m,n}(i_{\text{A3}}))$. For given $\alpha_n(i_{\text{A3}}), P_m(i_{\text{A3}})$, and $q_n(i_{\text{A3}})$, $\bm v_{m,n}(i_{\text{A3}}+1)$ can be obtained via (\ref{mrc2}) and \textbf{Algorithm 1}.  Since problem (\ref{sp3}) is convex, thus we have 
		\begin{equation} \label{apc1} 
		\begin{aligned}
&t_{\text{In}}(\alpha_n(i_{\text{A3}}), P_m(i_{\text{A3}}),  q_n(i_{\text{A3}}), \bm v_{m,n}(i_{\text{A3}}+1))\ge \\
&t_{\text{In}}(\alpha_n(i_{\text{A3}}), P_m(i_{\text{A3}}),  q_n(i_{\text{A3}}), \bm v_{m,n}(i_{\text{A3}})).
		\end{aligned}
	\end{equation} 

Then, for given $ q_n(i_{\text{A3}})$ and $ \bm v_{m,n}(i_{\text{A3}}+1)$, $\alpha_n(i_{\text{A3}}+1)$ and  $P_m(i_{\text{A3}}+1)$ can be obtained
based on (\ref{21a}) and (\ref{21b}). In view of the linearity of (\ref{p3d}), we have 
		\begin{equation} \label{apc2} 
	\begin{aligned}
		&t_{\text{In}}(\alpha_n(i_{\text{A3}}+1), P_m(i_{\text{A3}}+1),  q_n(i_{\text{A3}}), \bm v_{m,n}(i_{\text{A3}}+1))\ge \\
		&t_{\text{In}}(\alpha_n(i_{\text{A3}}), P_m(i_{\text{A3}}),  q_n(i_{\text{A3}}), \bm v_{m,n}(i_{\text{A3}})).
	\end{aligned}
\end{equation} 

Next, for given $\alpha_n(i_{\text{A3}}+1), P_m(i_{\text{A3}}+1)$, and  $\bm v_{m,n}(i_{\text{A3}}+1)$, $q_n(i_{\text{A3}}+1)$ can be obtained via \textbf{Algorithm 2}. Since problem (\ref{rp3b})  is also convex, thus we have 
		\begin{equation} \label{apc3} 
	\begin{aligned}
		&t_{\text{In}}(\alpha_n(i_{\text{A3}}+1), P_m(i_{\text{A3}}+1),  q_n(i_{\text{A3}}+1), \bm v_{m,n}(i_{\text{A3}}+1))\ge \\
		&t_{\text{In}}(\alpha_n(i_{\text{A3}}), P_m(i_{\text{A3}}),  q_n(i_{\text{A3}}), \bm v_{m,n}(i_{\text{A3}})).
	\end{aligned}
\end{equation} 

			\begin{table*}
	\setcounter{equation}{40}
	\begin{equation} \label{mmse} 	
\begin{aligned}
			(\bm v_{m,n}^{\text {B}})^*\triangleq   \dfrac{\left( \bm h_{m,n}(\bm h_{m,n})^{\text {H}}+\sum\limits_{j=1, j\neq n}\bm h_{m,j}(\bm h_{m,j})^{\text {H}}+\frac{\sigma_w^2}{P_m}\bm I_K \right)^{-1}\bm h_{m,n} }{\left\|\left( \bm h_{m,n}(\bm h_{m,n})^{\text {H}}+\sum\limits_{j=1, j\neq n}\bm h_{m,j}(\bm h_{m,j})^{\text {H}}+\frac{\sigma_w^2}{P_m}\bm I_K \right)^{-1}\bm h_{m,n}\right\|}.
		\end{aligned}
	\end{equation}
	\hrule 
\end{table*}

Additionally, given that $\alpha_n$ and $q_n$ fall within the bounds of $\left[ 0, 1\right]$, and $P_m$ is constrained by $P_{\max}$, it is evident that the objective value of problem (\ref{p1}) possesses a finite upper bound. Consequently, this guarantees that the proposed algorithm will converge to a stable point, ensuring at minimum a sub-optimal solution for problem (\ref{p1}).

\section{Benchmark algorithms}

\subsection{The algorithm with TDMA}
\subsubsection{Problem Formulation}This problem can be obtained from the formulated problem by replacing $\text{Pr}(n)$ with $\tau_n$, where $\tau_n$ denotes the transmission time of the $n$-th BD. More specifically, this redefined optimization problem is articulated as
	\setcounter{equation}{36}
	\begin{equation} \label{tdma} 
	\begin{split}	
		& \underset{\beta_m, P_m, \alpha_n, \tau_n, \bm v_{m,n}}{\mathop{\max}}\,  \underset{\forall n}{\mathop{\min}}\, \tau_n\sum\limits_{m=1}^M \beta_m r_{m,n}^{\text {BD}}\\
		&\text{s.t.}~C_1, C_4, C_5, C_6, C_7,\\
		&\quad ~~C_{2}^{\text{TD}}: \sum\limits_{n=1}^N \tau_n=1, \\
		&\quad ~~C_{3}^{\text{TD}}: \sum\limits_{n=1}^N\tau_nR_{n}^{\text {AD}} \ge R_{\min}.\\
	\end{split}
\end{equation} 

\subsubsection{Algorithm Design} This problem can be solved with the proposed algorithm, except for the optimization of $\tau_n$. Upon determining the other variables, we can optimize $\tau_n$ using linear programming techniques. The detailed process is omitted for brevity.

\subsection{The algorithm with TC}
\subsubsection{Problem Formulation} 
Under this scheme, the achievable throughput of the AD can be rewritten as 
	\begin{equation} \label{fd1} \small
	\begin{aligned}	
		R_N^{\text {AD}}{=}\sum\limits_{m=1}^M\beta_m \log_2\left( 1{+}\frac{|(\bm v_{m}^{\text {A}})^{\text {H}}\bm h_m^{\text {d}}|^2P_m}{\sum\limits_{n=1}^N \alpha_n|h_{m,n}^{\text {f}}|^2|(\bm v_{m}^{\text {A}})^{\text {H}}\bm h_n^{\text {b}}|^2P_m+\sigma_w^2}\right) .
	\end{aligned}
\end{equation}

And the achievable throughput of the $n$-th BD can be rewritten as 
	\begin{equation} \label{fd2} \small 
	\begin{aligned}	
	\!\!	R_n^{\text {BD}}{=}\!\sum\limits_{m=1}^M \!\beta_m \log_2\left(\! 1{+}\frac{\alpha_n|h_{m,n}^{\text {f}}|^2|(\bm v_{m,n}^{\text {B}})^{\text {H}}\bm h_n^{\text {b}}|^2P_m}{\sum\limits_{j=1, j\neq n}\alpha_j|h_{m,j}^{\text {f}}|^2|(\bm v_{m,n}^{\text {B}})^{\text {H}}\bm h_j^{\text {b}}|^2P_m+\sigma_w^2}\right) .
	\end{aligned}
\end{equation}

Similarly, the CAP of each BD is not in focus. Accordingly,  the corresponding  optimization problem can be expressed as 
\begin{equation} \label{fdma} 
	\begin{split}	
		& \underset{\beta_m, P_m, \alpha_n, \bm v_{m,n}^{\text {B}}, \bm v_m^{\text {A}}}{\mathop{\max}}\,   \underset{\forall n}{\mathop{\min}}\, R_n^{\text {BD}}\\
		&\text{s.t.}~C_1, C_4, C_5, C_6, C_7,\\
		&\quad ~~C_{3}^{\text{FD}}:  R_N^{\text {AD}} \ge R_{\min}.\\
	\end{split}
\end{equation} 

\subsubsection{Algorithm Design}

For the given problem, our algorithm remains applicable once we exclude the solution for the CAP sub-problem. Besides, in addressing the beamforming vector of the backscattering signal from the prior section, the MRC should be substituted with the minimum-mean-square-error (MMSE), which can be expressed as (\ref{mmse}), where $\bm h_{m,n}\triangleq \sqrt{\alpha_n}h_{m,n}^{\text {f}} \bm h_n^{\text {b}}$. The detailed process is omitted for brevity.


	\end{appendices}

\end{document}